\documentclass{jfm}
\usepackage{textcomp}
\usepackage{graphicx}
\usepackage{epstopdf, epsfig}
\usepackage{color}
\usepackage{amsmath}
\usepackage{amssymb}
\usepackage[numbers]{natbib}
\usepackage{fancyhdr}
\usepackage[dvipsnames]{xcolor}
\usepackage[normalem]{ulem}

\newcommand\cites[1]{\citeauthor{#1}'s\ (\citeyear{#1})}

\DeclareMathOperator{\ReRe}{\mathrm{Re}}

\shorttitle{Statistical zonostrophic instability with magnetic field}
\shortauthor{C. Wang, J. Mason and A. D. Gilbert}

\title{On statistical zonostrophic instability and the effect of magnetic fields}

\author{Chen Wang\aff{1,2,3}\corresp{\email{wangchen@uic.edu.cn}}, 
  Joanne Mason\aff{3}  \and  Andrew D. Gilbert\aff{3}}

\affiliation{\aff{1}Research Center of Mathematics, Advanced Institute of Natural Sciences, Beijing Normal University, Zhuhai 519087, PR China

\aff{2}Department of Mathematical Sciences,  BNU-HKBU United International College,  Zhuhai 519087, PR China

\aff{3}Department of Mathematics and Statistics, University of Exeter,
Exeter,  EX4 4QF,  UK\\}

\begin{document}

\maketitle

\abstract

Zonal flows are mean flows in the east--west direction, which are ubiquitous on planets,  and can be formed through `zonostrophic instability': within turbulence or random waves, a weak large-scale zonal flow can grow exponentially to become prominent.  In this paper, we study the  statistical behaviour of the zonostrophic instability and the effect of magnetic fields. We use a stochastic white noise forcing to drive random waves, and study the growth of a mean flow in this random system. The dispersion relation for the growth rate of the expectation of the mean flow is derived, and properties of the instability are discussed. In the limits of weak and strong magnetic diffusivity,  the dispersion relation reduces to manageable   expressions, which provide clear insights into the effect of the magnetic field and scaling laws for the threshold 
of instability. The magnetic field mainly plays a stabilising role and thus impedes the formation of the zonal flow, but under certain conditions it can also have destabilising effects.  Numerical simulation of the stochastic flow is performed to confirm the theory. Results indicate that the magnetic field can significantly increase the randomness of the zonal flow. It is found that  the zonal flow of an individual realisation may behave very differently from the expectation. For weak magnetic diffusivity and moderate magnetic field strengths, this leads to considerable variation of the outcome, that is whether zonostrophic instability takes place or not in  individual realisations.

\section{Introduction}
Zonal flows are mean flows in the east--west direction, and are commonly found on Earth and other planets. They are perhaps most prominent on Jupiter, where belts of strong zonal jets are the main visible feature of its surface.  Numerous studies have been undertaken to understand the properties of zonal flows.  A review of this topic can be found in the recent book \citet{Galperin19}; here we summarise representative literature most relevant to our study.

 \citet{Rhines75} identified that zonal jets have a scale associated with the wavenumber $k_\beta= \sqrt{\beta/2U}$, where $U$ is the root-mean square (rms) zonal velocity and $\beta$ is the gradient of the Coriolis parameter. This scale is now known as the Rhines scale. Above this length scale, the inverse cascade of turbulence is suppressed by the $\beta$-effect, and turbulence  transfers energy to the zonal jets. \citet{Williams78} performed numerical simulations that reproduce
the zonal flows on both the Earth and Jupiter. From the conditions under which zonal flows emerge, he concluded that the $\beta$-effect and forcing are the two elements essential for the formation of zonal jets.  \citet{Vallis93} identified the important scales of turbulence in the zonal jets, and also showed that in addition to the $\beta$-effect, topography can also generate zonal flows. 
 \citet{Smith04} and \citet{Scott07} considered quasi-geostrophic flows and found that the effect of introducing a finite deformation radius is to increase the level of $\beta$ required for zonal jets to form. When the deformation radius is small enough, zonal jets are confined near the equator. 
\citet{Galperin06} studied  the energy spectrum of turbulent flows with strong zonal jets, and found that they exhibit a $-5$ power law, in what they termed a `zonostrophic regime', and in contrast to the celebrated Kolmogorov $-\frac{5}{3}$ power law of isotropic turbulence.  
\citet{Dritschel08} proposed a mechanism for the formation of zonal jets: the mixing of potential vorticity (PV) results in staircases in the PV profile.

 To further understand mechanisms for jet formation, \citet{Farrell03,Farrell07} established a compact system that combines the evolution of turbulence and mean flow. By studying the `structural stability'   of this system, they showed that the state with zero mean flow can be unstable, giving rise to the formation of zonal jets.  \citet{Srinivasan12} advanced the theory by undertaking an analytical study of the instability problem and deriving an explicit dispersion relation, also introducing the term `zonostrophic instability'.
 \citet{Parker13,Parker14}  incorporated weak nonlinearity and derived a Ginzburg--Landau equation for the zonal-flow amplitude, which they used to model the generation of zonal jets in terms of pattern formation.

In astrophysical contexts, it is necessary to consider the effect of magnetic fields on the flow and study the magnetohydrodynamic (MHD) problem.  \citet{Diamond05} have given a comprehensive review of zonal flows in plasmas, where the magnetic field plays a key role. Recent studies have also revealed the important link between zonal flows and cross helicity \citep{Heinonen23}, which is a central topic of MHD. Zonal flows may also be important for the structure of the solar tachocline, which is generally believed to be the source of the Sun's magnetic field (for a comprehensive review of the solar tachocline, see the monograph edited by \cite{Hughes07}).   In this vein,  \citet{Tobias07} and \citet{Tobias11} have undertaken numerical simulation of MHD beta-plane flow in Cartesian and spherical geometry. Their results indicate that even a weak magnetic field can suppress the formation of zonal jets, and this may explain the lack of observations so far of zonal jets in the Sun, where strong magnetic fields interact with plasma flows.  In terms of theory, \citet{Constantinou18} undertook an MHD zonostrophic stability analysis using the method of \citet{Srinivasan12}. Their results confirmed that the magnetic field indeed reduces the growth rate of the zonostrophic instability. \citet{Durston16} and \citet{Algatheem23} considered MHD zonostrophic instability for large-scale shear flows and for deterministic Kolmogorov flow, respectively. Analytical approximations can been made by exploiting scale separation between large-scale flow and small-scale waves, and these indicate that the magnetic field can inhibit the hydrodynamic zonostrophic instability, but it can also generate a new branch of unstable modes for such shear flows.

While it appears that a magnetic field suppresses  zonostrophic instability and zonal flows,  the physical mechanism for the suppression requires further exploration.  To this end,   \citet{Constantinou18} and \citet{Constantinou19} considered magnetic field subject to the tilting present in shear flows.   They showed that the Maxwell stress imposed by the magnetic field opposes the Reynolds stress,   and therefore inhibits the formation of zonal flows.   \citet{Chen20} and \citet{Chen21} considered turbulent magnetic fields are strong and highly disordered. Via averaging the system over an intermediate scale,  they showed that the resulting Maxwell stress again inhibits the growth of any zonal flow. 

In the present paper, we will study the zonostrophic instability with an emphasis on its statistical properties and the effect of magnetic fields. In particular, we consider the growth of weak mean flows in a system of random waves driven by white noise. We will derive a dispersion relation for exponential growth of the expectation of the mean flow, and analyse its properties in detail. We will also compare the theory with numerical simulations of the stochastic system, to assess its validity and the assumptions made.  Our method for the instability analysis differs from previous studies which are mainly based on spatio-temporal correlation functions, e.g. \citet{Farrell03,Farrell07}, \cite{Srinivasan12} and \cite{Constantinou18}. 
The use of  spatio-temporal correlation functions works well for hydrodynamic zonostrophic instability, but becomes complicated when a magnetic field is added.  For example, \cite{Constantinou18} describe the calculations leading to the dispersion relation for the MHD zonostrophic instability as `complicated and unilluminating', and they do not document its full expression. This is mainly because the presence of the magnetic field introduces several new terms that are quadratic in the fields: for example, one has to include the correlation between the velocity and the magnetic field to close the system. Our study, on the other hand, directly analyses random waves and avoids such complication. Our derivation only involves temporal correlation of the random waves,
which significantly simplifies the mathematics. We are able derive a  compact dispersion relation in certain limiting cases of the parameters, which provides simple scaling laws for the condition of instability.

 An outcome of our simplified dispersion relation is a clear explanation for the mechanism by which the magnetic field affects zonostrophic instability.  The physical explanation of \citet{Constantinou18,  Constantinou19,  Chen20,  Chen21} is based on the form of the magnetic terms in the mean-flow equation. Our  interpretation, on the other hand, is based on the dispersion relation for  zonostrophic instability, in which the effect of the magnetic field on the growth of the mean flow is more evident.  Also, we do not assume \textit{a priori}  the time and length scales of the flow but solve the full mathematical problem under the quasi-linear approximation, allowing a comprehensive survey of parameter space.  With this, we observe further  effects that can arise in the presence of a magnetic field in addition to the Maxwell stress discussed by previous authors: under certain conditions, the magnetic field can  change the effective  viscosity of the flow, or affect the instability through the interaction between the mean flow and the mean field. 

In many previous studies of zonostrophic instability, an ergodic assumption has been used:   
namely that the zonal mean velocity in each realisation is the same as in  the ensemble average, in other words the zonal flow itself  has no stochasticity. 
This ergodic assumption has been widely used for analyses of mean flows in various problems, for example, zonostrophic instability \citep{Srinivasan12, Constantinou18}, the structural stability of turbulent jets \citep{Farrell03, Farrell07},  pattern formation of zonal flows \citep{Parker13, Parker14,Ruiz16}, and numerical simulation of turbulent zonal flows \citep{Marston08, Tobias07, Tobias11}.  Its validity has not been established in general, though \cite{Bouchet13} provide a systematic argument based on separation of time-scales between fluctuating and mean fields, making use of a Fokker--Planck formulation.
In the present study, we will use numerical simulations to examine the ergodic assumption in the context of  zonostrophic instability with and without magnetic field. We run multiple simulations to see if individual realisations of the mean flow behave similarly.

Another issue that arises in the MHD zonostrophic instability concerns modes with complex growth rates. \citet{Constantinou18} showed that in the presence of magnetic field, unstable zonostrophic modes can possess complex growth rates, a phenomenon uncommon for pure hydrodynamic zonostrophic instability \citep[e.g.][]{Srinivasan12,Ruiz16}.  Although the real part of a complex growth rate can be positive, the system's  evolution remains  poorly understood in this case, since these modes lie in the parameter regime where \citet{Tobias07} reported no zonal flow formation. We will undertake numerical simulations to reveal the behaviour of these complex modes, paying special attention to the stochastic nature of the flow.

The organisation of the paper is as follows: in \S 2, we perform an analysis of hydrodynamic zonostrophic instability. Although the hydrodynamic problem is relatively well understood, we use it as a basic example 
   to establish our method of analysis, and lay the foundation for the MHD problem. We present the governing equations and the quasi-linear approximation in \S2.1.  
Then we establish the `basic state' in \S 2.2, which is a forced--dissipative system without mean flow.  
The stability of this state is analysed in \S 2.3, and a dispersion relation is derived for the growth rate of the zonal mean flow. Its properties are briefly discussed in \S 2.4. Finally in \S 2.5, numerical simulations are undertaken to verify the theory. The analysis for the MHD zonostrophic instability is undertaken in  \S 3 following the same methodology, but it is a more complicated and interesting problem. In particular, the full dispersion relation derived in \S 3.3 is a rather unwieldy expression; however we derive simplified dispersion relations in \S 3.4 in certain asymptotic limits of the parameters. In the discussion of the dispersion relation in \S 3.5, we give a comprehensive investigation of the effect of the magnetic field strength and  magnetic diffusivity, and compare this with previous studies. In  numerical simulations in \S 3.6, we demonstrate the strong impact of the magnetic field on the stochasticity of the evolving zonal flows. Concluding remarks are given in \S 4.  

\section{Hydrodynamic zonostrophic instability}

In this section, we present the analysis for the hydrodynamic zonostrophic instability. We give a straightforward derivation of a dispersion relation for the unstable growth rate of the zonal flow; this takes a relatively simple form and is equivalent to that obtained by \citet{Srinivasan12}. We do this to set out the methodology for the magnetic zonostrophic instability  that will be studied in  the next section.

\subsection{Governing equations}
We start with the 2D vorticity equation on the $\beta$ plane:
\begin{equation}
\zeta_t-\psi_y\zeta_x+\psi_x\zeta_y+\beta\psi_x=F-\mu\zeta+\nu \nabla^2 \zeta,\qquad \zeta=\nabla^2\psi\,. \label{2.1}
\end{equation}
Here the $x$- and $y$-directions are the longitudinal and latitudinal directions, respectively, $\psi$ is the stream function,  $\zeta$ is the vorticity,  $F$ is an external forcing which we will specify later, $\mu$ is the bottom  drag and $\nu$ is the viscosity. The variables have been non-dimensionalised by characteristic length and time scales.   
 
We then decompose the flow as
\begin{equation}
  \psi=\overline{\psi}(y,t)+\psi'(x,y,t),\quad \zeta=\overline{\zeta}(y,t)+\zeta'(x,y,t)\,, \label{2.2} 
\end{equation}
where the overline represents the zonal average:
\begin{equation}
  \overline{\psi}=\frac{k}{2\pi}\int_0^{{2\pi}/{k}}\psi\, \mathrm{d}x,\quad \overline{\zeta}=\frac{k}{2\pi}\int_0^{{2\pi}/{k}}\zeta\, \mathrm{d}x,
\end{equation}
and $\psi'$ and $\zeta'$ are the fluctuating fields.  Now we apply the quasi-linear approximation: substituting (\ref{2.2}) into (\ref{2.1}), we neglect all the nonlinear terms except those with mean components. Using $U=-\overline{\psi}_y$  to denote the mean zonal velocity, we derive the equation for the fluctuating vorticity:
\begin{equation}
\zeta'_t+U\zeta'_x+(\beta-U_{yy})\psi_x'=F-\mu\zeta'+\nu\nabla^2\zeta',\qquad \zeta'=\nabla^2\psi'. \label{2.4}
\end{equation}
To find the evolution equation for $U$, we take the zonal average of (\ref{2.1}) and find
\begin{equation}
  U_t-\left(\overline{\psi'_x\psi'_y}\right)_y=-\mu U+\nu U_{yy}, \label{2.5}
\end{equation}
taking the forcing $F$ to have zero zonal average. 

We consider the forcing
\begin{equation}\label{2.6}
  F=\sigma \hat{\xi}(t)e ^{\mathrm{i}kx}+\mathrm{c.c.},
\end{equation}
where $\sigma$ is the strength of the forcing, $k$ is the wavenumber in the $x$-direction, $\hat{\xi}$ is a complex Gaussian white noise, and `c.c.' represents the complex conjugate of the previous term (or terms).    The expectation of the white noise has the properties
\refstepcounter{equation}\label{noise}
$$
   \mathbb{E}\bigl[\hat{\xi}(t_1)\hat{\xi}^*(t_2)\bigr]=\delta(t_1-t_2), \qquad  \mathbb{E}\bigl[\hat{\xi}(t_1)\hat{\xi}(t_2)\bigr]=0,\qquad \mathbb{E}\bigl[\hat{\xi}(t)\bigr]=0,  \eqno(\theequation a, b, c)
$$
which indicates that the values of the white noise at two different times are independent (\ref{noise}$a$); it has zero expectation (\ref{noise}$c$), and its statistical properties are independent of time. Because of these  properties,  white noise as forcing is a standard idealisation for stochastic differential equations. In our problem, we use it to drive waves with random amplitudes, as a very idealised model of turbulence. More details of the white noise and our numerical implementation are given in Appendix A.

The forcing (\ref{2.6}) is the same as that used by \citet{Farrell03,Farrell07}. It also 
has  similarities with the approach of \citet{Srinivasan12}, who use a `ring forcing', namely the stochastic driving of a ring of wavenumbers of given radius in Fourier space (for details, see Appendix B). The ring forcing is essentially isotropic, so the $\beta$-effect is necessary for the zonostrophic instability. This property of isotropy may be more relevant to modelling 
the formation of zonal jets on planets. Our forcing (\ref{2.6}), on the other hand, has a single wavenumber, and is therefore a `point forcing' which is anisotropic. Although it may not be as realistic as the isotropic ring forcing, it can still reveal key properties of zonostrophic instability. For MHD instabilities, because of the complexity of the analysis, we simplify to a point forcing as a first step, and leave any elaboration to an isotropic ring forcing for future research.

\subsection{The `basic state'}

We now consider the stability problem governed by equations (\ref{2.4}) and (\ref{2.5}). Because of the white-noise forcing, $\psi$ is stochastic, which
complicates typical stability analyses.
The usual method to remove stochasticity is to use a spatio-temporal correlation function  which will evolve deterministically \citep{Farrell03,Farrell07,Srinivasan12}. We take a different approach: we directly solve for $\psi$ in terms of the white noise and so retain its randomness. When we proceed to the equation for the mean flow, where quadratic terms of the fundamental waves will appear, we compute the expectation, taking advantage of the properties of the white noise given in equations (\ref{noise}$a,b,c$).

The forcing $F$ in (\ref{2.6}) is independent of $y$ and generates waves. We consider a state in which there is zero zonal mean flow, $U$, as the  `basic state', upon which zonostrophic instability develops. We thus take a solution of the form
\begin{equation}\label{2.15}
 \psi'=\psi_1=\hat{\psi}_1(t)e^{\mathrm{i}kx}+\mathrm{c.c.},\quad U=0.
\end{equation}
Substitution into (\ref{2.4}) yields
\begin{equation}\label{2.16}
  \frac{\mathrm{d}\hat{\psi}_1}{\mathrm{d}t}+\left(-\mathrm{i}\, \frac{\beta}{k}+\mu+\nu k^2\right)\hat{\psi_1}=-\frac{\sigma }{k^2}\, \hat{\xi}.
\end{equation}
For simplicity, we  assume that $\hat{\psi}_1=0$ at $t=0$. The solution for $\hat{\psi}_1$ is then
\begin{equation}\label{2.17}
  \hat{\psi}_1=-\frac{\sigma}{k^2}\, e^{\lambda_1 t}\int_0^t\hat{\xi}(\tau) e^{-\lambda_1\tau }\mathrm{d}\tau,
\end{equation}
where
\begin{equation}\label{2.18}
  \lambda_1=-\mu-\nu k^2+\mathrm{i}\, \frac{\beta}{k}
\end{equation}
is the  eigenvalue of the homogeneous system (\ref{2.15}).  The fluctuating flow $\hat{\psi}_1$ has the form of a damped Rossby wave, with its amplitude driven by the white noise. It is unsteady and stochastic,  but statistically, $\hat{\psi}_1$ has zero expectation and its probability density will settle down to a steady distribution as $t\rightarrow \infty$, as can be checked by solving the corresponding Fokker--Planck equation. 

\subsection{Instability analysis }

We then study the stability of the basic state in the statistical sense, adding now perturbations $\psi_2$ and $U_2$:
\begin{equation}\label{2.19}
  \psi'=\psi_1+\psi_2 + \cdots, \quad U=U_2 + \cdots .
\end{equation}
Assuming that $\psi_2$ and $U_2$ are small, we substitute (\ref{2.19}) into (\ref{2.4}) and (\ref{2.5}), and then linearise 
terms involving
$\psi_2$ and $U_2$, which gives
\begin{equation}
  \zeta_{2,t}+\beta\psi_{2,x}+\mu\zeta_2-\nu\nabla^2\zeta_2=-U_2\zeta_{1,x}+U_{2,yy}\psi_{1,x}\qquad \zeta_2=\nabla^2\psi_2, \label{2.20}
\end{equation}
\begin{equation}
  U_{2,t}+\mu U_2-\nu U_{2,yy}=\left(\overline{\psi_{1,y}\psi_{2,x}+\psi_{2,y}\psi_{1,x}}\right)_y.  \label{2.21}
\end{equation}
We seek solutions in the form 
\begin{equation}\label{2.22}
  U_2=\hat{U}(t) e^{\mathrm{i}my}+\mathrm{c.c.}, \quad \psi_2=\hat{\psi}_{21}(t) e^{\mathrm{i}kx+\mathrm{i}my}+\hat{\psi}_{22}(t) e^{-\mathrm{i}kx+\mathrm{i}my}+\mathrm{c.c.},
\end{equation}
that is, the mean flow $U_2$ has a wavenumber $m$ in the transverse direction, and $\psi_2$ has a wavenumber combination of the basic wave and the mean flow. Substituting into (\ref{2.20}) and (\ref{2.21}), we obtain
\begin{equation}\label{2.23}
  \frac{\mathrm{d}\hat{\psi}_{21}}{\mathrm{d} t}-\lambda_2\hat{\psi}_{21}=-\mathrm{i}k \Lambda \hat{U}\hat{\psi}_1,
\end{equation}
\begin{equation}\label{2.24}
  \frac{\mathrm{d}\hat{\psi}_{22}}{\mathrm{d} t}-\lambda_2^*\hat{\psi}_{22}=\mathrm{i}k \Lambda \hat{U}\hat{\psi}_1^*,
\end{equation}
\begin{equation}\label{2.25}
  \frac{\mathrm{d}\hat{U}}{\mathrm{d}t}+(\mu+m^2\nu)\hat{U}=\mathrm{i}m^2k(\hat{\psi}_{21}\hat{\psi}_1^*-\hat{\psi}_{22}\hat{\psi}_1),
\end{equation}
where
\begin{equation}\label{2.26}
 \lambda_2=-\mu-\nu(k^2+m^2)+\frac{\mathrm{i}k\beta}{k^2+m^2}\, ,\qquad \Lambda= \frac{k^2-m^2}{k^2+m^2}\, .
\end{equation}
We again take $\hat{\psi}_{21}$ and $\hat{\psi}_{22}$ to be zero at $t=0$. The solution of (\ref{2.23}) and (\ref{2.24}) is then
\begin{equation}\label{2.27}
  \hat{\psi}_{21}=-\mathrm{i}k\Lambda e^{\lambda_2 t}\int_{0}^t\hat{U}(\tau)\hat{\psi}_1(\tau) e^{-\lambda_2 \tau}\mathrm{d}\tau,
\end{equation}
\begin{equation}\label{2.28}
  \hat{\psi}_{22}=\mathrm{i}k\Lambda e^{\lambda_2^* t}\int_0^t\hat{U}(\tau)\hat{\psi}_1^*(\tau) e^{-\lambda_2^* \tau}\mathrm{d}\tau.
\end{equation}

We substitute (\ref{2.27}) and (\ref{2.28}) into (\ref{2.25}), and then obtain an equation for $\hat{U}$:
\begin{align}\label{2.29}
  &\frac{\mathrm{d}\hat{U}}{\mathrm{d}t}+(\mu+\nu m^2)\hat{U} \nonumber \\
  &=m^2k^2\Lambda\int_0^t\hat{U}(\tau)\left[\hat{\psi}_1(\tau)\hat{\psi}_1^*(t) e^{-\lambda_2(\tau-t)}+\hat{\psi}_1^*(\tau)\hat{\psi}_1(t)e^{-\lambda_2^*(\tau-t)}\right]\mathrm{d}\tau.
\end{align}
This equation describes the evolution of the mean flow driven by Rossby waves. 
The quantity $\hat{U}$ is again stochastic, but our interest is its expectation $\mathbb{E}(\hat{U})$. In particular, we look for the solution that grows exponentially, to indicate zonostrophic instability. Thus, we consider
\begin{equation}\label{2.33}
  \mathbb{E}[\hat{U}(t)]=\tilde{U}e^{s t},
\end{equation}
and the main objective of the analysis is to solve for  the growth rate $s$.

In order to proceed mathematically, we need to make the assumption that the mean flow $\hat{U}(\tau)$ and the quadratic term of the fundamental wave $\hat{\psi}_1^*(\tau)\hat{\psi}_1(t)$ are statistically uncorrelated, hence their expectations are separable: 
\begin{equation}\label{assumption}
  \mathbb{E}\bigr[\hat{U}(\tau)\hat{\psi}_1(\tau)\hat{\psi}_1^*(t)\bigr]=\mathbb{E}[\hat{U}(\tau)] \, \mathbb{E}\bigl[\hat{\psi}_1(\tau)\hat{\psi}_1^*(t)\bigr], 
\end{equation}
thus the expectation of (\ref{2.29}) becomes
\begin{align}\label{2.30}
  &\frac{\mathrm{d}\mathbb{E}(\hat{U})}{\mathrm{d}t}+(\mu+\nu m^2)\mathbb{E}(\hat{U}) \nonumber \\
  &=m^2k^2\Lambda\int_0^t\mathbb{E}[\hat{U}(\tau)]\left\{\mathbb{E}[\hat{\psi}_1(\tau)\hat{\psi}_1^*(t)] e^{-\lambda_2(\tau-t)}+\mathbb{E}[\hat{\psi}_1^*(\tau)\hat{\psi}_1(t)]e^{-\lambda_2^*(\tau-t)}\right\}\mathrm{d}\tau.
\end{align}
We do not have a proof for (\ref{assumption}): it is an assumption that we will make to derive an analytical dispersion relation, which we will test through comparison with numerical simulations.  But (\ref{assumption}) is directly related to the assumption of zonal-mean ergodicity that has been widely used in mean-flow dynamics \citep[e.g.][]{Srinivasan12, Farrell03, Marston08}. This assumption states that the zonal-mean velocity of an individual realisation is equal to the ensemble average, or expectation:
\begin{equation}\label{ergo}
  \hat{U}=\mathbb{E}[\hat{U}] ; 
\end{equation}
in this case since $\hat{U}$ is no longer stochastic, (\ref{ergo}) implies assumption (\ref{assumption}). Indeed, we will show that our result of zonostrophic instability is consistent with the result of \citet{Srinivasan12} based on the ergodic assumption. But unlike (\ref{ergo}), our assumption (\ref{assumption}) still retains stochasticity in the mean flow and is therefore a weaker assumption. We will refer to (\ref{ergo}) as the `full ergodic assumption' and (\ref{assumption}) as the `partial ergodic assumption' in what follows.

 The expectation of the fundamental-wave term $\hat{\psi}_1^*(\tau)\hat{\psi}_1(t)$ in (\ref{2.30}) may be computed explicitly: applying property (\ref{noise}$a$) to (\ref{2.17}), we find that for $t>\tau$,  
\begin{align} \label{2.31}
\mathbb{E}\bigl[\hat{\psi}_1(\tau)\hat{\psi}_1^*(t)\bigr]=&\frac{\sigma^2}{k^4}\, e^{\lambda_1^*t+\lambda_1\tau}\int_{p=0}^t\int_{q=0}^\tau\mathbb{E}[\hat{\xi}^*(p)\hat{\xi}(q)]e^{-\lambda_1^*p-\lambda_1q}\, \mathrm{d}q\, \mathrm{d}p  \nonumber \\
=& \frac{\sigma^2}{k^4}\, e^{\lambda_1^*t+\lambda_1\tau}\int_{p=0}^t\int_{q=0}^\tau\delta(p-q)e^{-\lambda_1^*p-\lambda_1q}\, \mathrm{d}q\, \mathrm{d}p \simeq-\frac{\sigma^2}{k^4}\frac{1}{\lambda_1+\lambda_1^*} \, e^{\lambda_1^*(t-\tau)}. \nonumber \\
\end{align}
In the last step, we have ignored the term arising from the initial condition, which is exponentially small as $t\to\infty$.

Substituting (\ref{2.33}) and (\ref{2.31}) into (\ref{2.30}), and again neglecting the exponentially small initial value term,  we obtain the dispersion relation determining the growth rate $s$:
\begin{equation}\label{2.34}
  s+\mu+\nu m^2=-\frac{m^2\sigma^2\Lambda}{k^2(\lambda_1+\lambda_1^*)}\left(\frac{1}{s-\lambda_1^*-\lambda_2}+\frac{1}{s-\lambda_1-\lambda_2^*}\right).
\end{equation}
We rewrite the dispersion relation in terms of the original variables:
\begin{equation}\label{2.35}
  s+\mu+\nu m^2=\frac{m^2\sigma^2(k^2-m^2)}{2 k^2(\mu +\nu k^2)}\frac{1}{(s+2\mu+2\nu k^2+\nu m^2)(k^2+m^2)+\mathrm{i}\beta m^2/k} +\mathrm{c.c.e.}s.\,,
\end{equation}
where the expression `c.c.e.\textit{s}.' represents the previous terms with all quantities complex conjugated \emph{except} $s$ (cf.\ equation (\ref{2.34})); this notation is helpful to give succinct equations in this paper, and we will use it  repeatedly in the MHD case.

To reduce the number of independent quantities, we rescale according to:
\begin{align}\label{2.36}
 & s=s_\star \sigma ^{\frac{2}{3}}, \quad m=m_\star k \quad \mu=\mu_\star \sigma^{\frac{2}{3}},\quad \nu=\nu_\star \sigma^{\frac{2}{3}}k^{-2},\quad \beta=\beta_\star k \sigma^{\frac{2}{3}}\,.
\end{align}
This corresponds to a non-dimensionalisation based on the forcing strength $\sigma$, the viscosity $\nu$, and the scale $k^{-1} $ of the forcing. The non-dimensional quantity obtained from the viscosity may be linked to a Grashof number given by $\mathrm{Gr} = \nu_\star^{-2}$ and $\beta_\star$ is a non-dimensional measure of the strength of the $\beta$-effect on the same basis \citep[see][]{Childress01,Durston16}.
Under this rescaling the dispersion relation (\ref{2.35}) becomes
 \begin{equation}\label{2.37}
   s_\star+\mu_\star+\nu_\star m_\star^2=\frac{m_\star^2(1-m_\star^2)}{2(\mu_\star+\nu_\star)}\frac{1}{(s_\star+2\mu_\star+2\nu_\star+\nu_\star m_\star^2)(1+m_\star^2)+\mathrm{i}\beta_\star m_\star^2} +\mathrm{c.c.e.}s.
 \end{equation}
The rescaled parameters will be convenient for finding  conditions for instability in the parameter space, and for deriving asymptotic expressions for the dispersion relation of MHD instabilities shown later on. However, when presenting general results, we will  mainly use the original parameters which are more relevant to the physics.

\subsection{Results and discussion}

We now briefly discuss properties of the dispersion relation (\ref{2.35}) or (\ref{2.37}), which governs hydrodynamic zonostrophic instability. First, we observe that our dispersion relation would exactly result from the analysis in \citet{Srinivasan12}, if we apply their derivation to our forcing. Although their discussion focused on a ring forcing, their equation (C16) is  a dispersion relation for general forms of forcing, and in our case results in (\ref{2.35}). We give details in Appendix B, and note that the agreement is not trivial, since we followed a very different derivation.

\begin{figure}
  \centering
  \includegraphics[width=0.45\linewidth]{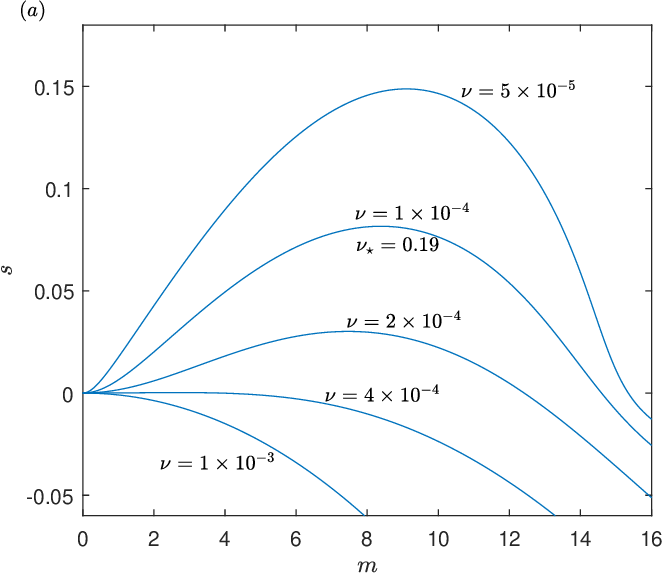}
  \includegraphics[width=0.51\linewidth]{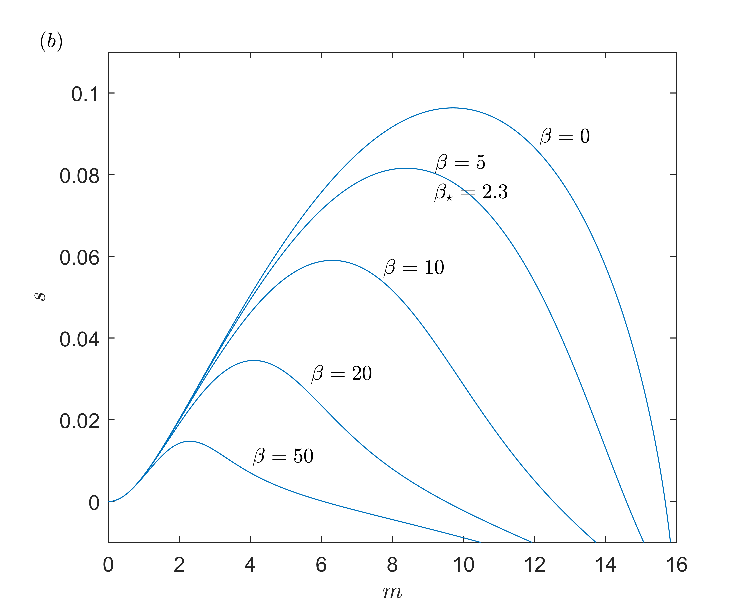}
  \caption{Dispersion relation for the hydrodynamic instability with $k=16$, $\mu=0$, $\sigma=0.05$. Panel $(a)$ plots the growth rates at different $\nu$ with $\beta=5$, and panel $(b)$ plots the growth rates at different $\beta$ with $\nu=10^{-4}$.   }\label{F1}
\end{figure}

Focusing on geophysical applications, \citet{Srinivasan12} studied the effects of the drag coefficient $\mu$ in detail, and mostly set the viscosity $\nu$ to be zero. Our present study focuses on astrophysical applications, and we will mainly consider the situation where the drag $\mu$ is zero and study the effect of the viscosity $\nu$. We will choose $k=16$ for the scale of the forcing, and $\sigma=0.05$ as the amplitude of the forcing. \citet{Tobias07} considered 
wavenumbers with $14<k<20$, which includes our $k=16$. The small amplitude $\sigma=0.05$ is intended to generate weak turbulence where the quasi-linear approximation is expected to be valid \citep[cf.][]{Srinivasan12}, and it is also of the same order as that used by \citet{Constantinou18}.  

 In figure \ref{F1}, we show the (real) growth rate $s$ versus vertical wavenumber $m$ at different $\nu$ and $\beta$ in panels $(a)$ and $(b)$ respectively, calculated from the dispersion relation (\ref{2.35}).  One feature of the dispersion relation is that as the drag coefficient $\mu$ is zero the growth rate $s$ approaches zero as $m\rightarrow0$. This contrasts with the case  $\mu>0$ studied by \citet{Srinivasan12}, where $s$ is negative at $m=0$. As seen in figure \ref{F1}($a$), increasing the viscosity reduces the growth rate and finally suppresses instability, as one would expect. In figure \ref{F1}($b$), we see that increasing $\beta$ also reduces instability \citep{Srinivasan12}, but the growth rates at smaller wavenumbers are unaffected, as is evident from the dispersion relation (\ref{2.35}). Note that zonostrophic instability continues to exist even for $\beta=0$. Even though there is now no background vorticity gradient present to give a preferred direction in the system, the basic state of the fluid system remains anisotropic because of the forcing we employ in (\ref{2.6}). As is known,  zonostrophic instability cannot occur in fluid systems that are isotropically forced, without some further mechanism to break symmetry, such as a $\beta$-effect or magnetic field \citep{Srinivasan12, Bakas13}
  
  From figure \ref{F1}, we see that neutral stability $s\to0$ occurs in the limit $m\rightarrow 0$. Applying $m\rightarrow 0$ to the dispersion relation (\ref{2.35}) or (\ref{2.37}) with $\mu=0$, we find that the condition for zonostrophic instability is
 \begin{equation}\label{2.38}
   \nu_\star<2^{-\frac{1}{3}}\qquad \mathrm{or}\qquad \nu<2^{-\frac{1}{3}}k\sigma ^{\frac{2}{3}} ;
 \end{equation}
the forcing must be strong enough to overcome the effect of viscous dissipation. As we just discussed, $\beta$ is absent from this condition, and we stress that (\ref{2.38}) only applies to the case $\mu_\star=0$, otherwise the instability threshold takes place at a finite $m$, which involves $\beta$ in the stability condition \citep[cf.][]{Srinivasan12}.
 In the rest of the paper, we will mainly pay attention to the situation where $\mu_\star\ll\nu_\star\ll1$ so that (\ref{2.38}) is relevant, while $\beta_\star$ is of order of unity or larger.

We comment on another interesting relation between our dispersion relation (\ref{2.37}) and that of \cite{Srinivasan12}: both these dispersion relations have the properties that  all unstable modes have real $s_\star$ and instability only exists for $m_\star<1$, despite the difference in the spatial structure of the forcing. \cite{Srinivasan12} indicated that they could not show these two important properties analytically in their case of ring forcing. However we can make some progress with our simpler dispersion relation: for real $s_\star$, we need the right-hand side of  (\ref{2.37}) to be positive so that $s_\star>0$ on the left-hand side is possible, and this requires $m_\star<1$. We have not yet been able to explain why $s_\star$ is always real when $\ReRe s_\star>0$. In any case this appears not to be a generic property: \citet{Ruiz16} have shown that for forcing with other spatial structures, unstable modes with complex growth rates indeed exist for hydrodynamic zonostrophic instability.  

 \subsection{Numerical simulation}
 
To verify our theory for hydrodynamic zonostrophic instability, we undertake numerical simulations for the flow governed by equation (\ref{2.1}), with further details in appendix A. We use a pseudospectral method with a spatial discretization of 256$\times$256 mesh points in the domain $[0, 2\pi/k]\times [0, 2\pi/m]$.
For temporal discretization, we use the Crank--Nicolson method, with the nonlinear terms advanced using Euler's method. 
It\^o's interpretation is used for integration of the white noise. A very small  time step of $\Delta t=0.005$ is used for both the temporal evolution and discretization for the white noise, 
to ensure a good approximation to the decorrelation property in (\ref{noise}$a$).
We choose the parameters $\beta=5$, $\nu=10^{-4}$, $\mu=0$, $\sigma=0.05$,  $k=16$ and $m=5$, corresponding to $\beta_\star=2.3$, $\nu_\star=0.19$, which have been used in figure \ref{F1}. We take $\psi=10^{-4}\cos(my)$ at $t=0$ to render a very weak zonal flow $U=-\overline{\partial_y\psi}$ initially.  Our main objective here is to verify the theory rather than aim for physical realism, hence we only consider one Fourier mode for the initial zonal flow to make comparison straightforward.     

\begin{figure}
  \centering
  \includegraphics[width=1\linewidth]{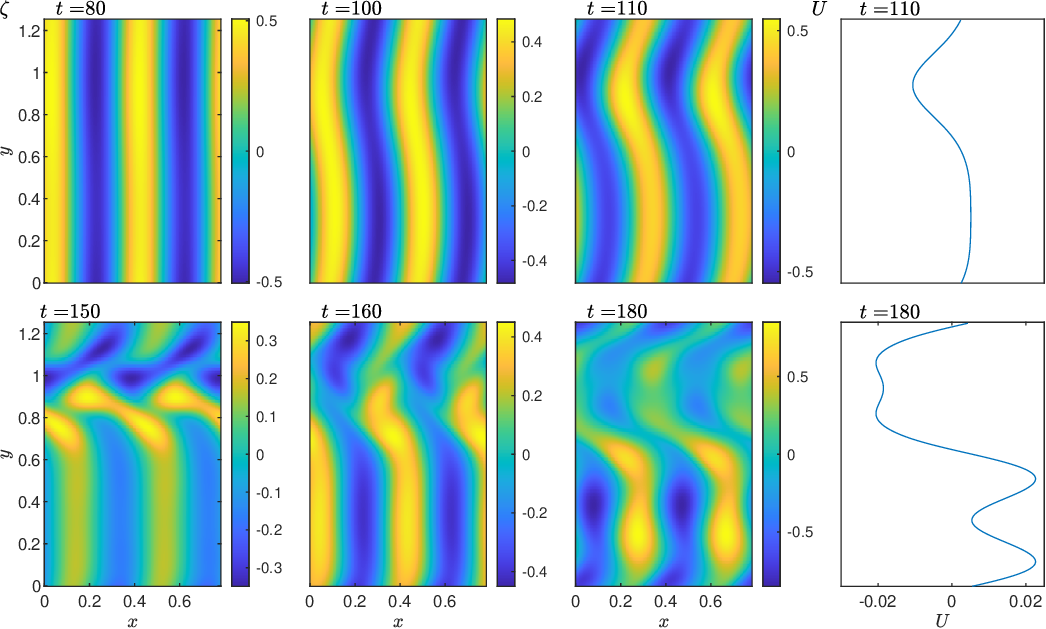}
  \caption{Numerical solution for the evolution of vorticity and zonal velocity without magnetic field, $\beta=5$, $k=16$, $\nu=10^{-4}$, $\mu=0$, $\sigma=0.05$, $m=5$, $\beta_\star=2.3$, $\nu_\star=0.19$, $m_\star=0.31$. }\label{F2}
\end{figure}

In figure \ref{F2},  we show an example of the evolution of the vorticity field, alongside two snapshots of the zonal mean flow profiles. At earlier times $t\leq 80$, the vorticity field has the same pattern as the forcing $F$ (i.e., it is periodic in the $x$-direction and homogeneous in the $y$-direction). Then unstable zonal flows gradually grow stronger, causing bending in the $y$-direction visible at $t=100$. At $t=110$, we observe that nonlinear effects become significant, generating a localised zonal jet near $y=1$. The jet continues to grow stronger, and then saturates and undergoes some complicated nonlinear evolution. At $t=180$, another jet in the opposite direction is visible around $y=0.2$.  
Note that a simulation with a different white noise results in a different flow pattern, but with qualitatively the same features as that in figure \ref{F2}.
 
\begin{figure}
  \centering
    \includegraphics[width=0.534\linewidth]{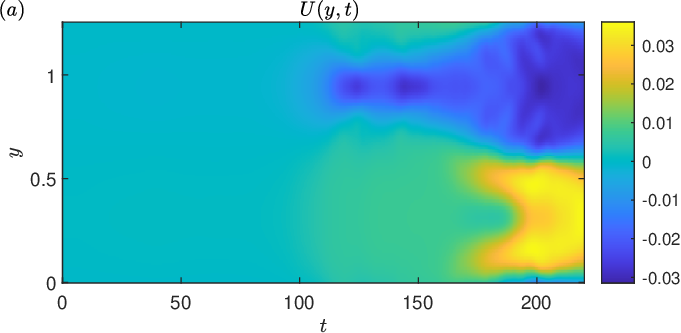}
        \includegraphics[width=0.458\linewidth]{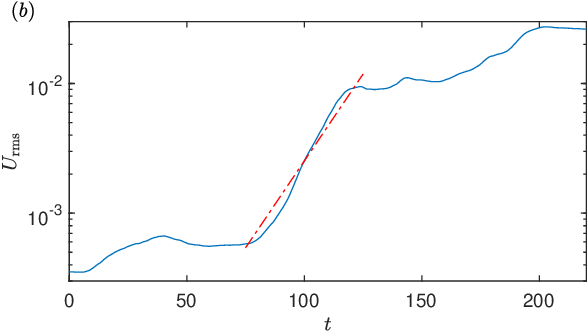}
  \caption{Evolution of the zonal velocity $U$ for figure \ref{F2}. Panel $(a)$ is a Hovm\"{o}ller diagram showing $U(y,t)$, and panel $(b)$ shows the rms value of $U$. The solid line is the solution of the numerical simulation, and the straight dash--dot line is the exponential growth predicted by the dispersion relation (\ref{2.35}), with growth rate $s=6.17\times 10^{-2}$.}     
  \label{F3}
\end{figure}

In figure \ref{F3},  we plot the evolution of the zonal flow; figure \ref{F3}$(a)$ shows the density field of $U(y,t)$ in a Hovm\"{o}ller diagram, giving the evolution of the mean flow with time on the horizontal axis (snapshots of $U(y,t)$ at $t=110$ and $t=180$ are shown in figure \ref{F2}).
The formation of the two jets is clearly seen.  Note that unlike previous studies on zonal jets (e.g. \citet{Srinivasan12}, \citet{Parker13}), we do not see jet-merging in our simulation. We think this may be due to the simple structure of forcing that we used to drive the waves (cf. (\ref{2.6})).

 In figure \ref{F3}$(b)$,  we show the rms velocity with respect to $y$,  i.e.
\begin{equation}
  U_\mathrm{rms}= \biggl(   {\frac{m}{2\pi}\int_0^{{2\pi}/{m}}U^2\,\mathrm{d}y}  \biggr)^{1/2}. \label{rms}
\end{equation}
We also plot the prediction of zonostrophic instability in a dash--dot line,   showing good agreement between the theory and the simulation for $t\in[80,120]$ after some transient behaviour up to $t\simeq 80$. The good agreement justifies the theory, including the quasi-linear approximation and the assumption of wave-mean uncorrelation (\ref{assumption}) used in the analysis.

\begin{figure}
  \centering
    \includegraphics[width=0.49\linewidth]{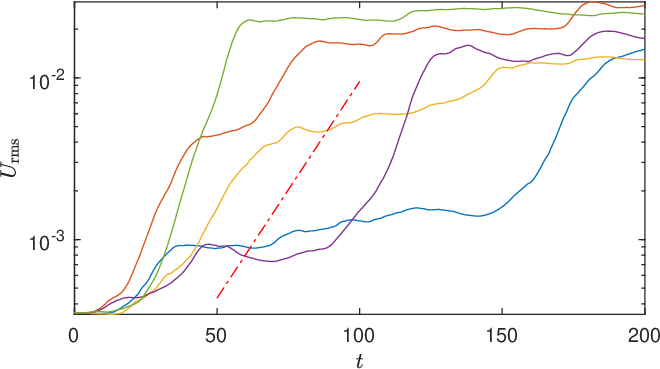}
  \caption{Five realisations of $U_\mathrm{rms}$ with different white-noise forcing; the parameters are the same as in figure \ref{F2}.  The straight dash--dot  line represents the exponential growth predicted by the dispersion relation (\ref{2.35}). 
 }\label{F4}
\end{figure}
To show the effect of different realisations of the forcing,  in figure \ref{F4} we plot the evolution of $U_\mathrm{rms}$ for five different examples of the driving white noise.  The exponential growth of zonostrophic instability theory is plotted in the dash--dot line. $U_\mathrm{rms}$ starts to grow at different times for different realisations,  but their growth rates are all very close to the theoretical prediction.
 
The simulation results allow us to make more comments on the full assumption of mean-flow ergodicity in (\ref{ergo}), that is assuming the mean flow has no stochasticity. Our simulations show that this assumption is clearly not satisfied, since $U_\mathrm{rms}$ is different for different realisations. But if our focus is on whether  zonostrophic instability occurs or not, then this assumption seems to be reasonable, because all realisations have periods of growth at  a similar rate as our prediction for the  expectation, derived under the weaker partial assumption (\ref{assumption}). The underlying reason for this property, i.e. that the mean flow exhibits similar growth rates  in different realisations,  
however, remains elusive. Previous theoretical studies have indicated that full mean-flow ergodicity holds when the flow reaches a statistically steady state \citep{Marston08}, when the drag $\mu$ is sufficiently strong \citep{Bouchet13}, or when there is time- or space-scale separation between the mean flow and the waves \citep{Bouchet13,Durston16}. None of these conditions applies to our case: the flow is in a transient state during the instability, our simulations used zero drag $\mu=0$, and our mean flow and waves have similar length scales. More theoretical work is required to understand when full or partial mean-flow ergodicity holds.

\section{MHD zonostrophic instability}
\subsection{Governing equations}
Following the same methodology, we study the MHD zonostrophic instability: we add a constant magnetic field $B_0$ in the $x$-direction and study its effect.   The original MHD equations for the 2D flow are
\begin{align}
 \zeta_t-\psi_y\zeta_x+\psi_x\zeta_y+\beta\psi_x & =-a_y j_x+a_x j_y+F-\mu\zeta+\nu \nabla^2 \zeta, \label{3.1}
\\
   a_t-\psi_y a_x+\psi_x a_y & =\eta \nabla^2 a, \label{3.2}
  \\
 \zeta  =\nabla^2\psi,\;\;   j & =\nabla^2 a\,. \label{3.1a}
 \end{align}
Here $a$ is the magnetic potential, i.e. the magnetic field is $\textbf{\textit{B}}=(-a_y, a_x, 0)$, and $j$ is the current density. We again apply the quasi-linear approximation for this system. For the flow, we apply the same decomposition as (\ref{2.2}). For the magnetic field, we decompose the wave and mean by
\begin{equation}
  a=\overline{a}(y,t)+a'(x,y,t) + \cdots ,\quad  j=\overline{j}(y,t)+j'(x,y,t)  + \cdots , \label{3.3}
\end{equation}
and for the mean field we set
\begin{equation}
  -\overline{a}_y=B_0+B(y,t), \label{3.4}
\end{equation}
where $B_0$ is the externally added constant mean field in the $x$-direction, while  $B$ is a small variation of magnetic field averaged in the $x$-direction and caused by the flow.   
We assume $|B_0|\gg |B|$. We substitute (\ref{3.3}) and (\ref{3.4}) into (\ref{3.1}) and (\ref{3.2}), and then apply the same quasi-linear approximation: we only keep the nonlinear terms involving the flow velocity $U$ and field $B$. The governing equations for the waves are
\begin{equation}
  \zeta'_t+U\zeta'_x+(\beta-U_{yy})\psi'_x-(B_0+B)j'_x+B_{yy} a'_x=-\mu\zeta'+\nu \nabla^2 \zeta'+F, \label{3.5}
\end{equation}
\begin{equation}
  a'_t+U a'_x-(B_0+B)\psi'_x=\eta \nabla^2 a', \label{3.6}
\end{equation}
and the mean flow and field evolve according to
\begin{equation}
  U_t-\overline{(\psi_x'\psi_y'-a_x'a_y')_y}=-\mu U+\nu U_{yy}, \label{3.7}
\end{equation}
\begin{equation}
  B_t-\overline{(a'\psi'_x)}_{yy}=\eta B_{yy}. \label{3.8}
\end{equation}
The term $a_x'a_y'$ is often referred to as the Maxwell stress, which has the opposite sign but  a similar structure to the Reynolds stress $\psi_x'\psi_y'$. Intuitively, one might expect that the Maxwell stress is the mechanism by which a magnetic field can inhibit zonal flows, but we need to solve the actual zonostrophic instability problem to reach a concrete conclusion.

\subsection{Basic state}
For the basic state, the forcing $F$ generates waves with zero  mean flow and perturbation field, $U=B=0$. With the forcing (\ref{2.6}), we consider  solutions of the form 
\begin{equation}\label{3.9}
  \psi'=\psi_1=\hat{\psi}_1(t)e^{\mathrm{i}kx}+\mathrm{c.c.},\quad a'=a_1=\hat{a}_1(t)e^{\mathrm{i}kx}+\mathrm{c.c.},
\end{equation}
and we seek the amplitudes $\hat{\psi}_1(t)$ and $\hat{a}_1(t)$.
Substituting (\ref{2.6}) and (\ref{3.9}) into (\ref{3.5}) and (\ref{3.6}) yields two ODEs in $t$, which we write in vector form:
\begin{equation}\label{3.10}
  \frac{\mathrm{d}}{\mathrm{d}t}\left(\begin{array}{ll}\hat{\psi}_1\\ [8pt]\hat{a}_1\end{array}\right)=\left(\begin{array}{ll}\displaystyle{\frac{\mathrm{i}\beta}{k}}-\mu-\nu k^2\quad& \mathrm{i}kB_0\\ [8pt]\mathrm{i}kB_0\quad&-\eta k^2\end{array}\right)\left(\begin{array}{ll}\hat{\psi}_1\\ [8pt]\hat{a}_1\end{array}\right)+\left(\begin{array}{ll}\displaystyle{-\frac{\sigma}{k^2}}\hat{\xi}\\ [8pt]0\end{array}\right).
\end{equation}
Their solutions are
\begin{align}\label{3.11}
  \hat{\psi}_1& = -\Psi_+e^{\lambda_{1+}t}\int_0^t e^{-\lambda_{1+}r}\hat{\xi}(r)\mathrm{d}r
    +\Psi_-e^{\lambda_{1-}t}\int_0^te^{-\lambda_{1-}r}\hat{\xi}(r)\mathrm{d}r,
\\
\label{3.12}
  \hat{a}_1 & =-Ae^{\lambda_{1+}t}\int_0^t e^{-\lambda_{1+}r}\hat{\xi}(r)\mathrm{d}r+A e^{\lambda_{1-}t}\int_0^te^{-\lambda_{1-}r}\hat{\xi}(r)\mathrm{d}r,
\end{align}
where
\begin{eqnarray}\label{3.13}
&\Psi_{\pm}=\displaystyle\frac{\sigma}{2k^2 Q}\Bigl(\frac{\mathrm{i}\beta}{k}-\mu-\nu k^2+\eta k^2 \pm Q\Bigr),\quad A= \displaystyle\frac{\mathrm{i}\sigma B_0}{kQ}, \nonumber\\
& \displaystyle \lambda_{1\pm}=\frac{1}{2}\Bigl(\frac{\mathrm{i}\beta}{k}-\mu-\nu k^2-\eta k^2\pm Q\Bigr),\quad Q=\Bigl[\Bigl(\displaystyle\frac{\mathrm{i}\beta}{k}-\mu-\nu k^2+\eta k^2\Bigr)^2-4k^2B_0^2\Bigr]^\frac{1}{2}.\nonumber \\
\end{eqnarray}
Note that $\lambda_{1\pm}$ are the two eigenvalues of the matrix in (\ref{3.10}). In the limit $B_0\rightarrow 0$, 
\begin{equation}
  \lambda_{1+}\rightarrow \frac{\mathrm{i}\beta}{k}-\mu-\nu k^2=\lambda_1,\quad \lambda_{1-}\rightarrow -\eta k^2. 
\end{equation}
Hence, $\lambda_{1+}$ recovers the hydrodynamic eigenvalue of Rossby waves and $\lambda_{1-}$ is the rate of Ohmic damping of a  magnetic field mode.

\subsection{Instability analysis}

Upon the basic state, we add the disturbances of zonostrophic instability, denoted by a subscript `2':
\begin{equation}\label{3.15}
  \psi'=\psi_1+\psi_2 + \cdots  ,\quad a'=a_1+a_2 + \cdots ,\quad U=U_2 + \cdots ,\quad B=B_2 + \cdots .
\end{equation}
Substituting  (\ref{3.15}) into (\ref{3.5}--\ref{3.8}) and then linearising the `2' components, we obtain for the fluctuating fields, 
\begin{align}\label{3.16}
  \zeta_{2,t}+\beta\psi_{2,x}-B_0 \nabla^2 a_{2,x}+\mu \zeta_2-\nu \nabla^2 \zeta_2 
&  =-U_2 \zeta_{1,x}+U_{2,yy}\psi_{1,x}+B_2 \nabla^2 a_{1,x}-B_{2,yy}a_{1,x},\\
\label{3.17}
  a_{2,t}-B_0\psi_{2,x}-\eta \nabla^2 a_2 & =-U_2 a_{1,x}+B_2 \psi_{1,x},  \\
   \zeta_2 & =\nabla^2 \psi_2\,,
\end{align}
and for the mean fields, 
\begin{align}\label{3.18}
  U_{2,t}+\mu U_2-\nu U_{2,yy} & =\overline{\psi_{1,x}\psi_{2,yy}+\psi_{2,x}\psi_{1,yy}-a_{1,x}a_{2,yy}-a_{2,x}a_{1,yy}}\,,
\\
\label{3.19}
  B_{2,t}-\eta B_{2,yy} & =\left(\overline{a_1\psi_{2,x}+a_2\psi_{1,x}}\right)_{yy}.
\end{align}
Note that here we are referring to $B_2$ as the `mean' (magnetic) field for succinctness, and will continue to do so, though really it is only the perturbation component of the full mean field $B_0 + B_2(y,t) + \cdots $, with the mean always taken in $x$. 

Similar to the hydrodynamic case, we seek solutions in the form 
\begin{align}\label{3.20}
  U_2 & =\hat{U}(t) e^{\mathrm{i}my}+\mathrm{c.c.}, \quad \psi_2=\hat{\psi}_{21}(t) e^{\mathrm{i}kx+\mathrm{i}my}+\hat{\psi}_{22}(t) e^{-\mathrm{i}kx+\mathrm{i}my}+\mathrm{c.c.} , \nonumber \\
  B_2 & =\hat{B}(t) e^{\mathrm{i}my}+\mathrm{c.c.}, \quad a_2=\hat{a}_{21}(t) e^{\mathrm{i}kx+\mathrm{i}my}+\hat{a}_{22}(t) e^{-\mathrm{i}kx+\mathrm{i}my}+\mathrm{c.c.}
\end{align}
Substituting (\ref{3.20}) into (\ref{3.16}--\ref{3.19}), we obtain the evolution equations for the fluctuating amplitudes, 
\begin{align}\label{3.21}
  \frac{\mathrm{d}}{\mathrm{d}t}\left(\begin{array}{ll}\hat{\psi}_{21}\nonumber \\ [5pt] \hat{a}_{21}\end{array}\right)=&\left(\begin{array}{ll}\displaystyle\frac{\beta\mathrm{i}k}{m^2+k^2}-\mu-\nu(m^2+k^2)\quad & \mathrm{i}kB_0\nonumber \\ [8pt] \mathrm{i}kB_0\quad & -\eta(m^2+k^2)\end{array}\right)
  \left(\begin{array}{ll}\hat{\psi}_{21}\nonumber \\ [5pt] \hat{a}_{21}\end{array}\right)\\
 & -\mathrm{i}k  \left(\begin{array}{ll}\Lambda (\hat{\psi}_1\hat{U}-\hat{a}_1\hat{B}) \\ [5pt] \hat{a}_{1}\hat{U}-\hat{\psi}_1\hat{B}\end{array}\right),
\end{align}
\begin{align}\label{3.22}
  \frac{\mathrm{d}}{\mathrm{d}t}\left(\begin{array}{ll}\hat{\psi}_{22}\nonumber \\ [5pt] \hat{a}_{22}\end{array}\right)=&\left(\begin{array}{ll}\displaystyle-\frac{\beta\mathrm{i}k}{m^2+k^2}-\mu-\nu(m^2+k^2)\quad & -\mathrm{i}kB_0\nonumber \\ [8pt] -\mathrm{i}kB_0\quad & -\eta(m^2+k^2)\end{array}\right)
  \left(\begin{array}{ll}\hat{\psi}_{22}\nonumber \\ [5pt] \hat{a}_{22}\end{array}\right)\\
 & +\mathrm{i}k  \left(\begin{array}{ll}\Lambda (\hat{\psi}_1^*\hat{U}-\hat{a}_1^*\hat{B}) \\ [5pt] \hat{a}_{1}^*\hat{U}-\hat{\psi}_1^*\hat{B}\end{array}\right),
\end{align}
and for the mean flow and field,
\begin{align}\label{3.23}
\frac{\mathrm{d}\hat{U}}{\mathrm{d}t}+\mu \hat{U}+\nu m^2\hat{U} &=\mathrm{i}k m^2\bigl(\hat{\psi}_1^*\hat{\psi}_{21}-\hat{a}_1^*\hat{a}_{21}\bigr)-\mathrm{i}km^2\bigl(\hat{\psi}_1\hat{\psi}_{22}-\hat{a}_1\hat{a}_{22}\bigr),
\\
\label{3.24}
  \frac{\mathrm{d}\hat{B}}{\mathrm{d}t}+\eta m^2\hat{B} &=\mathrm{i}km^2\bigl(\hat{\psi}_1^*\hat{a}_{21}-\hat{a}_1^*\hat{\psi}_{21}\bigr)-\mathrm{i}km^2\bigl(\hat{\psi}_1\hat{a}_{22}-\hat{a}_1\hat{\psi}_{22}\bigr).
\end{align}
The solutions of (\ref{3.21})  are
\begin{align}\label{3.25}
\hat{\psi}_{21}& =\int_0^t\Bigl\{\Bigl[-\Lambda D_+\hat{\psi}_1(\tau)+M\hat{a}_1(\tau)\Bigr]\hat{U}(\tau)-\Bigl[M\hat{\psi}_1(\tau)-\Lambda D_+\hat{a}_1(\tau)\Bigr]\hat{B}(\tau)\Bigr\}e^{\lambda_{2+}(t-\tau)}\mathrm{d}\tau\nonumber \\
& +\int_0^t\Bigl\{\Bigl[\Lambda D_-\hat{\psi}_1(\tau)-M\hat{a}_1(\tau)\Bigr]\hat{U}(\tau)+\Bigl[M\hat{\psi}_1(\tau)-\Lambda D_-\hat{a}_1(\tau)\Bigr]\hat{B}(\tau)\Bigr\}e^{\lambda_{2-}(t-\tau)}\mathrm{d}\tau,
\\
\label{3.26}
  \hat{a}_{21}& =\int_0^t\Bigl\{\Bigl[\Lambda M\hat{\psi}_1(\tau)+D_-\hat{a}_1(\tau)\Bigr]\hat{U}(\tau)-\Bigl[D_-\hat{\psi}_1(\tau)+\Lambda M\hat{a}_1(\tau)\Bigr]\hat{B}(\tau)\Bigr\}e^{\lambda_{2+}(t-\tau)}\mathrm{d}\tau  \nonumber\\
  & +\int_0^t\Bigl\{-\Bigl[\Lambda M\hat{\psi}_1(\tau)+D_+\hat{a}_1(\tau)\Bigr]\hat{U}(\tau)+\Bigl[D_+\hat{\psi}_1(\tau)+\Lambda M\hat{a}_1(\tau)\Bigr]\hat{B}(\tau)\Bigr\}e^{\lambda_{2-}(t-\tau)}\mathrm{d}\tau,
\end{align}
where
\begin{align}\label{3.27}
  \lambda_{2\pm}=\frac{\mathrm{i}\Omega_2-\mu-(\nu+\eta)k_2^2\pm Q_2}{2}\, ,\quad  \quad Q_2=\Bigl[\Bigl(\mathrm{i}\Omega_2-\mu-\nu k_2^2+\eta k_2^2\Bigr)^2-4k^2B_0^2\Bigr]^\frac{1}{2} , \nonumber \\
D_{\pm}=\frac{\mathrm{i}k(\mathrm{i}\Omega_2-\mu-\nu k_2^2+\eta k_2^2\pm Q_2)}{2Q_2}\, ,\quad M=\frac{k^2B_0}{Q_2}\, ,\quad   \Omega_2=\frac{\beta k}{m^2+k^2}\, ,\quad  k_2^2=m^2+k^2.
\end{align}
For the solution of (\ref{3.22}), we notice that the right-hand side of (\ref{3.21}) is the complex conjugate of the right-hand side of (\ref{3.22}), except $\hat{U}$ and $\hat{B}$ remain the same. Therefore, we may find the solutions for the `22' components by simply taking the complex conjugate of (\ref{3.25}) and (\ref{3.26}), and then replace all occurrences of $\hat{U}^*$ and $\hat{B}^*$ by $\hat{U}$ and $\hat{B}$.

We then substitute (\ref{3.25}) and (\ref{3.26}) into (\ref{3.23}) and (\ref{3.24}) to find a system of equations for the mean velocity and field
\begin{align}\label{3.28}
  &\frac{\mathrm{d}\hat{U}}{\mathrm{d}t}+\mu\hat{U}+\nu m^2\hat{U} & \nonumber \\
 &  =  \mathrm{i}m^2k\int_0^t\Bigl\{\Bigl[-\Lambda D_+\hat{\psi}_1^*(t)\hat{\psi}_1(\tau)+M\hat{\psi}_1^*(t)\hat{a}_1(\tau)-\Lambda M\hat a^*_1(t)\hat{\psi}_1(\tau)-D_-\hat{a}_1^*(t)\hat{a}_1(\tau)\Bigr]\hat{U}(\tau) \Bigr. \nonumber \\
& +\Bigl.\Bigl[-M \hat{\psi}_1^*(t)\hat{\psi}_1(\tau)+\Lambda D_+\hat{\psi}_1^*(t)\hat{a}_1(\tau)+D_-\hat{a}_1^*(t)\hat{\psi}_1(\tau)+\Lambda M \hat{a}_1^*(t)\hat{a}_1(\tau)\Bigr]\hat{B}(\tau) \Bigr\} e^{\lambda_{2+}(t-\tau)}\mathrm{d}\tau \nonumber \\
& +\mathrm{i}m^2k\int_0^t\Bigl\{\Bigl[\Lambda D_-\hat{\psi}_1^*(t)\hat{\psi}_1(\tau)-M\hat{\psi}_1^*(t)\hat{a}_1(\tau)+\Lambda M\hat a^*_1(t)\hat{\psi}_1(\tau)+D_+\hat{a}_1^*(t)\hat{a}_1(\tau)\Bigr]\hat{U}(\tau) \Bigr. \nonumber \\
& +\Bigl.\Bigl[M \hat{\psi}_1^*(t)\hat{\psi}_1(\tau)-\Lambda D_-\hat{\psi}_1^*(t)\hat{a}_1(\tau)-D_+\hat{a}_1^*(t)\hat{\psi}_1(\tau)-\Lambda M \hat{a}_1^*(t)\hat{a}_1(\tau)\Bigr]\hat{B}(\tau) \Bigr\} e^{\lambda_{2-}(t-\tau)}\mathrm{d}\tau  \nonumber \\
& + \mathrm{c.c.e.}\hat{U}.\hat{B}.,
\end{align}
\begin{align}\label{3.29}
&\frac{\mathrm{d}\hat{B}}{\mathrm{d}t}+\eta m^2\hat{B} \nonumber \\
&=\mathrm{i}m^2k\int_0^t\Bigl\{\Bigl[\Lambda M\hat{\psi}_1^*(t)\hat{\psi}_1(\tau)+D_-\hat{\psi}_1^*(t)\hat{a}_1(\tau)+\Lambda D_+\hat{a}^*_1(t)\hat{\psi}_1(\tau)-M\hat{a}_1^*(t)\hat{a}_1(\tau)\Bigr]\hat{U}(\tau)\Bigr. \nonumber \\
&-\Bigl.\Bigl[D_-\hat{\psi}^*_1(t)\hat{\psi}_1(\tau)+\Lambda M \hat{\psi}^*_1(t)\hat{a}_1(\tau)-M\hat{a}_1^*(t)\hat{\psi}_1(\tau)+\Lambda D_+\hat{a}^*_1(t)\hat{a}_1(\tau)\Bigr]\hat{B}(\tau)\Bigr\}e^{\lambda_{2+}(t-\tau)}\mathrm{d}\tau \nonumber \\
&+\mathrm{i}m^2k\int_0^t\Bigl\{\Bigl[-\Lambda M\hat{\psi}_1^*(t)\hat{\psi}_1(\tau)-D_+\hat{\psi}_1^*(t)\hat{a}_1(\tau)-\Lambda D_-\hat{a}^*_1(t)\hat{\psi}_1(\tau)+M\hat{a}_1^*(t)\hat{a}_1(\tau)\Bigr]\hat{U}(\tau)\Bigr. \nonumber \\
&+\Bigl.\Bigl[D_+\hat{\psi}^*_1(t)\hat{\psi}_1(\tau)+\Lambda M \hat{\psi}^*_1(t)\hat{a}_1(\tau)-M\hat{a}_1^*(t)\hat{\psi}_1(\tau)+\Lambda D_-\hat{a}^*_1(t)\hat{a}_1(\tau)\Bigr]\hat{B}(\tau)\Bigr\}e^{\lambda_{2-}(t-\tau)}\mathrm{d}\tau \nonumber \\
&+ \mathrm{c.c.e.}\hat{U}.\hat{B}.
\end{align}
The notation `c.c.e.$\hat{U}$.$\hat{B}$.' means the complex conjugate of the previous terms except $\hat{U}$ and $\hat{B}$ remain unchanged (cf. `c.c.e.$s$' for (\ref{2.34}) and (\ref{2.35})). There is a proliferation of terms in the MHD mean flow and field equations compared with the hydrodynamic case (cf. (\ref{2.29})), as we have doubled the number of fields and waves in the basic state ($\hat{\psi}_1$ and $\hat{a}_1$) and in the harmonics ($\psi_{21}$, $a_{21}$, $\psi_{22}$, $a_{22}$), and now have two mean fields ($\hat{U}$ and $\hat{B}$).

The next step is to take the expectation of (\ref{3.28}) and (\ref{3.29}), and to seek exponentially growing solutions for the mean flow and mean field expectations 
\begin{equation}\label{3.30}
    \mathbb{E}[\hat{U}(t)]=\tilde{U} e^{st}, \quad \mathbb{E}[\hat{B}(t)]=\tilde{B} e^{st}.
\end{equation}
As for hydrodynamic instability, we assume that the mean flow and the mean field are both statistically uncorrelated with the fundamental waves, leading to the separation of expectations following our partial ergodic assumption: 
\begin{align}\label{assumption2}
  &\mathbb{E}[\hat{U}(\tau)\hat{\psi}_1^*(t)\hat{\psi}_1(\tau)]= \mathbb{E}[\hat{U}(\tau)]\, \mathbb{E}[\hat{\psi}_1^*(t)\hat{\psi}_1(\tau)],
  & \mathbb{E}[\hat{B}(\tau)\hat{\psi}_1^*(t)\hat{a}_1(\tau)]= \mathbb{E}[\hat{B}(\tau)]\,   \mathbb{E}[\hat{\psi}_1^*(t)\hat{a}_1(\tau)], \nonumber \\
 \end{align}
and similarly for related terms. Equation (\ref{assumption2}) again is a little weaker than the corresponding full ergodic assumption, namely
\begin{equation}\label{ergo2}
 \hat{U}=  \mathbb{E}[\hat{U}],\quad  \hat{B}=  \mathbb{E}[\hat{B}],
\end{equation}
that both the mean flow and field have no stochasticity \citep{Constantinou18}. 
 
The expectations of the fundamental waves $\hat{\psi}_1^*(t)\hat{\psi}_1(\tau)$, $\hat{\psi}_1^*(t)\hat{a}_1(\tau)$ etc.\  can now be computed as in (\ref{2.31}), and then applying (\ref{3.30}) to  the expectation of (\ref{3.28}) and (\ref{3.29}) provides the dispersion relation, with details deferred to appendix C. The final result is
\begin{equation}\label{3.31}
  \left(\begin{array}{ll}
  (s+\mu+\nu m^2) \tilde{U}   \\
  (s+\eta m^2) \tilde{B}
  \end{array}\right)=\left( \begin{array}{ll}
  N_{UU} & N_{UB} \\
  N_{BU} & N_{BB}
  \end{array}\right) \left(\begin{array}{ll}
 \tilde{U}   \\
 \tilde{B}
  \end{array}\right)\,.
\end{equation}
The terms $N_{UU}$, $N_{UB}$, $N_{BU}$ and $N_{BB}$ are functions of $s$ and parameters, with expressions given in (\ref{Ns}), and have clear physical meaning: $N_{UB}$ represents the nonlinear feedback to the mean flow $\tilde{U}$ from the mean field $\tilde{B}$, etc. The condition for a non-trivial solution for $\tilde{U}$ and  $\tilde{B}$   yields the dispersion relation which determines  eigenvalues $s$:
\begin{equation}\label{3.32}
   (s+\mu+\nu m^2-N_{UU})(s+\eta m^2-N_{BB})=N_{UB}N_{BU}.
\end{equation}
When the basic state magnetic field is switched off,  i.e.\ $B_0=0$, the coupling terms $N_{UB}$ and $N_{BU}$ are zero, and the two eigenvalues are given by
\refstepcounter{equation}
$$
 \label{3.33}
  s+\mu+\nu m^2=N_{UU},\quad s+\eta m^2 =N_{BB}. \eqno{(\theequation a,b)}
$$
Expression (\ref{3.33}$a$) is the dispersion relation for the hydrodynamic zonostrophic instability (\ref{2.35}), and (\ref{3.33}$b$)  is
\begin{equation}\label{3.34}
  s+\eta m^2+\frac{m^2\sigma^2}{2k^2(\mu+\nu k^2)}\biggl(\frac{1}{s+\mu+\nu k^2+\eta(m^2+k^2)+\mathrm{i}\beta/k}+\mathrm{c.c.e.}s\biggr)=0\,.
\end{equation}
It always gives a negative growth rate, representing the damping  of the mean magnetic field by the flow. As in the hydrodynamic case, we undertake a rescaling to reduce the number of parameters by applying (\ref{2.36}) together with 
\begin{align}\label{3.35}
  & B_0=B_{0\star}\sigma^{\frac{2}{3}}k^{-1},\quad \eta=\eta_\star k^{-2}\sigma ^{\frac{2}{3}},
\end{align}
and then (\ref{3.32}) can be expressed as
\begin{equation}
  s_\star=s_\star(m_\star, \mu_\star, \nu_\star,  \beta_\star, B_{0\star}, \eta_\star). \label{3.36}
\end{equation}
It is difficult to compare our dispersion relation directly with that of \citet{Constantinou18} because, as they state, their expression was complicated and uninformative, and so not published in their paper. Their result was also for ring forcing, unlike ours for  point forcing. Nonetheless, we will see that our numerical solution of the dispersion relation bears some similarity to theirs. 

\subsection{Asymptotic dispersion relations in limiting cases}

The expression for the full dispersion relation, (\ref{3.32}) or (\ref{3.36}), is complicated, but in certain limits of the parameters, it is possible to obtain simpler expressions, as explained in appendix C, that can  provide deeper insights into the effect of the magnetic field and the mechanisms driving instability. These limits include strong magnetic diffusivity, and weak magnetic diffusivity with strong and weak magnetic field. When discussing the applicability of these expressions, we assume $\mu_\star\ll\nu_\star\ll1$, while $\beta_\star$ is of order of unity or larger, as previously.

\subsubsection{The limit of large $\eta_\star$}
In the limit of $\eta_\star\rightarrow \infty$, the dispersion relation approximates as
\begin{align} \label{3.39}
 &s_\star+\mu_\star+\nu_\star m_\star^2 \nonumber  \\
=&\frac{m_\star^2 (1-m_\star^2)}{2\left(\mu_\star+\nu_\star+\displaystyle\frac{B_{0\star}^2}{\eta_\star}\right)\left\{[s_\star+2\mu_\star+\nu_\star(m_\star^2+2)](1 +m_\star^2)+\displaystyle\frac{B_{0\star}^2}{\eta_\star}(2+m_\star^2)+\displaystyle\mathrm{i}\beta_\star m_\star^2\right\} } \nonumber\\
&+\mathrm{c.c.e.} s\,.
\end{align}
For finite $\eta_\star$, (\ref{3.39}) is valid when $\eta_\star \gg \beta_\star$ and $B_{0\star}^2/\eta_\star\ll1$. Equation (\ref{3.39}) may be interpreted as the hydrodynamic expression (\ref{2.34}) with $\lambda_1$ and $\lambda_{2}$ replaced by $\lambda_{1+}$ and $\lambda_{2+}$, which have asymptotic expressions
\begin{equation}
   \lambda_{1+}\sim \frac{\mathrm{i}\beta}{k}-\mu-\nu k^2-\frac{B_0^2}{\eta}\, ,\quad \lambda_{2+}\sim \frac{\mathrm{i}\beta k}{k^2+m^2}-\mu-\nu (m^2+k^2)-\frac{k^2 B_0^2}{\eta (m^2+k^2)}\label{lambda}
\end{equation}
in the large $\eta$ limit. Hence the magnetic field acts on the waves as though increasing the viscosity of the fluid, which has a stabilising effect. The solutions of $s_\star$ predicted by (\ref{3.39})  are real,  as in the hydrodynamic case.  As $B_{0\star}$ increases,  $s_\star$ will reduce and finally become negative,  so that the instability is suppressed by the field.  Equation (\ref{3.39}) remains a sound approximation in this limit, as we will demonstrate in \S 3.5 via comparison with numerical solutions of the full dispersion relation.  Note that the right-hand side of (\ref{3.39}) arises solely from the Reynolds stress,   which is strongly damped by the magnetic field;  the Maxwell stress itself is negligible  in this regime.
 
\subsubsection{The limit of small $\eta_\star$ and small $B_{0\star}$}

In the limit of $B_{0\star},\eta_\star\rightarrow 0$ while $B_{0\star}^2/\eta_\star$ remains finite,  the dispersion relation reduces to a different  expression  
 \begin{align} \label{3.41}
s_\star+\mu_\star+\nu_\star m_\star^2=&\frac{m_\star^2(1-m_\star^2)}{2(\mu_\star+\nu_\star)}\frac{1}{(s_\star+2\mu_\star+2\nu_\star+\nu_\star m_\star^2)(1+m_\star^2)+\mathrm{i}\beta_\star m_\star^2}-\frac{m_\star^2B_{0\star}^2}{2\eta_\star \beta_\star^2 s_\star}\nonumber \\
&+\mathrm{c.c.e.}s.
\end{align}
This expression corresponds to a simplified mean-flow evolution equation, 
 \begin{equation} \label{3.42}
  \frac{\mathrm{d}\hat{U}}{\mathrm{d}t}+\mu\hat{U}+\nu m^2\hat{U}=mk^2\int_0^t \bigl[\Lambda \hat{\psi}_1^*(t)\hat{\psi}_1(\tau)e^{\lambda_{2+}(t-\tau)}-\hat{a}_1^*(t)\hat{a}_1(\tau)e^{\lambda_{2-}(t-\tau)}+\mathrm{c.c.} \bigr]\hat{U}(\tau)\, \mathrm{d}\tau.
\end{equation}
The magnetic term in (\ref{3.41}) therefore comes from the quadratic term $\hat{a}_1^*(t)\hat{a}_1(\tau)$,  its negative sign exhibiting a stabilising effect of the magnetic field.
From the original mean flow equation (\ref{3.7}), we observe that $\hat{a}_1^*(t)\hat{a}_1(\tau)$ and $\hat{\psi}_1^*(t)\hat{\psi}_1(\tau)$ arise from the Maxwell and Reynolds stresses, respectively, and so the  stabilising effect of the magnetic field can be interpreted as the Maxwell stress counteracting the Reynolds stress.  The quadratic term from the magnetic field shares similarities with equation (39) of \citet{Chen20},  but our system incorporates more details of the time dependent waves, unlike in their system where the stochastic magnetic field is taken to be static.

For finite $\eta_\star$ and $B_{0\star}$,   (\ref{3.41}) is valid when $\eta_\star, B_{0\star}^2\ll \nu_\star$.    The magnetic term raises the order of the equation from cubic in (\ref{2.37}) to quartic, and with this we will shortly find some branches of complex solutions for $s_\star$.   The presence of complex eigenvalues $s_\star$ has important implications for the stochasticity of the flow, as we will see in \S 3.6 via numerical simulations. The expression  (\ref{3.41}) can be used to predict the field strength that makes $s_\star$ complex for all wavenumbers,  but not to predict the field that makes $\ReRe s_\star<0$ and thus suppresses the exponential growth.  Such magnetic field turns out to be very strong,  and we will demonstrate this regime subsequently in \S 3.4.3.

\subsubsection{The limit of  small $\eta_\star$ and large $B_{0\star}$ }

Finally we derive an asymptotic dispersion relation for relatively strong magnetic field and weak magnetic diffusivity.  In the limit of $\eta_\star\rightarrow 0$ and $B_{0\star}\rightarrow \infty$, we find
 \begin{equation}\label{3.44}
   (s_\star+\mu_\star+\nu_\star m_\star^2)s_\star=N_{UB\star}N_{BU\star},
 \end{equation}
 with
\begin{align}\label{3.45}
N_{UB\star}=&\frac{\mathrm{i}m_\star^2(m_\star^2+2)[\mathrm{i}\beta_\star-\nu_\star(m_\star^4+m_\star^2)]}{2B_{0\star}(\mu_\star+\nu_\star)(m_\star^2+1)[\nu_\star m_\star^4+(\mathrm{i}\beta_\star+2s_\star+2\mu_\star+3\nu_\star)m_\star^2+2(s_\star+\mu_\star+\nu_\star)]}\nonumber\\
&+\mathrm{c.c.e.}s,
\end{align} 
 \begin{align}
 N_{BU\star}=\frac{-\mathrm{i}m_\star^2[2\nu_\star (m_\star^4+2m_\star^2+1)+2\mu_\star(m_\star^2+1)+(2s_\star+\mathrm{i}\beta_\star )m_\star^2]}{2B_{0\star}(\mu_\star+\nu_\star)[\nu_\star m_\star^4+(\mathrm{i}\beta_\star+2s_\star+2\mu_\star+3\nu_\star)m_\star^2+2(s_\star+\mu_\star+\nu_\star)]}+\mathrm{c.c.e.}s. \label{3.46}
 \end{align}
Here $(N_{UB\star}, N_{BU\star})=\sigma ^{-\frac{2}{3}}(N_{UB}, N_{BU})$ (see (\ref{3.32})), while $N_{UU}$ and $N_{BB}$ may be neglected at leading order. The relation (\ref{3.44}) has some distinctive properties.  In the previous two asymptotic dispersion relations (\ref{3.39}) and (\ref{3.41}),  only $N_{UU}$ plays a role (corresponding also to (\ref{3.33})),   hence the hydrodynamic instability still has a major contribution,  although it is modified by the magnetic field. On the contrary,  in (\ref{3.44}),   $N_{UU}$ becomes negligible and $N_{UB}$ and  $N_{BU}$ are the leading-order terms.  This means the hydrodynamic instability is  suppressed by the strong magnetic field,  and the  interaction between the mean flow and mean field is the dominant effect,  which can yield zonostrophic instability.  Another distinguishing feature of  this dispersion relation  is that $\eta_\star$ does not appear in (\ref{3.44}--\ref{3.46}).  Hence when the magnetic diffusivity is weak enough,  it no longer affects the instability.

For finite $B_{0\star}$ and $\eta_\star$,  (\ref{3.44}) is valid when $B_{0\star}$ is at order $\nu_\star^{-1}$ or larger and $\eta_\star$ is much smaller than $\nu_\star$. The eigenvalue $s_\star$ in this regime is generally complex.  The relation (\ref{3.44}) can be used to predict the transition from $\ReRe s_\star >0$ to $\ReRe s_\star <0$ as $B_{0\star}$ increases,  and so  provide insights about the instability threshold.

 In the next section, we will compare the predictions of the asymptotic dispersion relations with  numerical solutions. We will also derive  scaling laws for   instability thresholds.

\subsection{Results and discussion}

We now discuss the solution for  the growth rate $s$ determined by the full  dispersion relation (\ref{3.32}), focussing on the effect of the magnetic field $B_0$ and the magnetic diffusivity $\eta$. We fix the other parameters at $\beta=5$, $\sigma=0.05$, $\nu=10^{-4}$, $\mu=0$, $k=16$ corresponding to $\beta_\star=2.3$, $\nu_\star=0.19$, as used previously for the hydrodynamic case.
\begin{figure}
  \centering
  \includegraphics[width=0.495\linewidth]{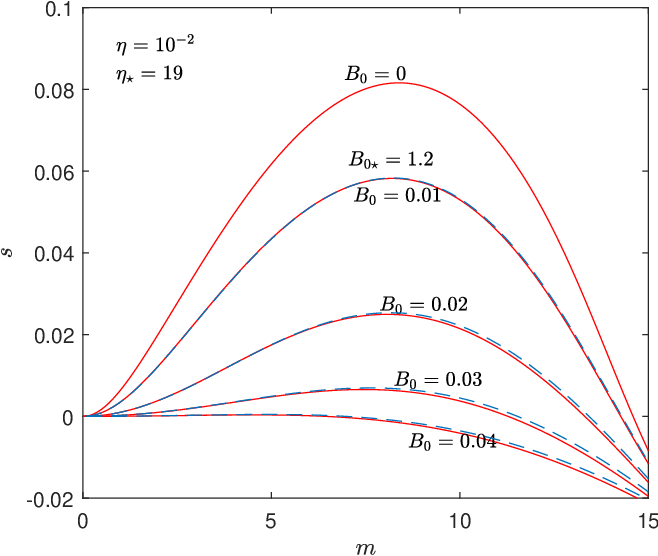}
  \caption{The dispersion relation giving $s$ for different magnetic field strengths $B_0$ at $\eta=10^{-2}$, corresponding to $\eta_\star=19$. The other parameters are $\beta=5$, $\sigma=0.05$, $\mu=10^{-4}$, $\mu=0$, $k=16$ corresponding to $\beta_\star=2.3$, $\nu_\star=0.19$.  Solid lines represent numerical solutions, and dashed lines represent the results of the asymptotic solution (\ref{3.39}).   }\label{F5}
\end{figure}

Figure \ref{F5} shows $s$ calculated from  the dispersion relation for various $B_0$ at a relatively high magnetic diffusivity $\eta=10^{-2}$,  corresponding to a large $\eta_\star=19$. The solid lines are the numerical solutions of (\ref{3.32}), and the dashed lines are the results of the asymptotic solution (\ref{3.39}) derived under the condition of large $\eta_\star$. All the solutions for the growth rate $s$ are real. The agreement between the asymptotic and numerical solutions is good for all of the cases presented. Increasing the magnetic field reduces the growth rate and finally suppresses the instability. As discussed above, the magnetic field tends to make the flow effectively more viscous, and so has a stabilising effect. 

\begin{figure}
  \centering
  \includegraphics[width=0.465\linewidth]{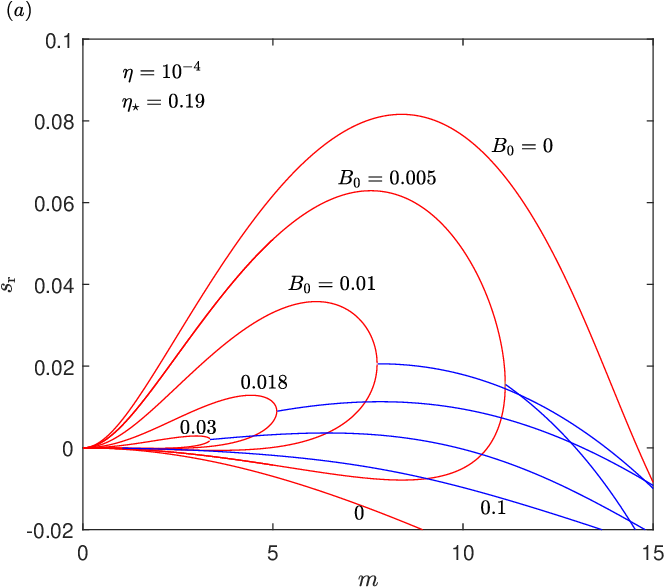}
    \includegraphics[width=0.52\linewidth]{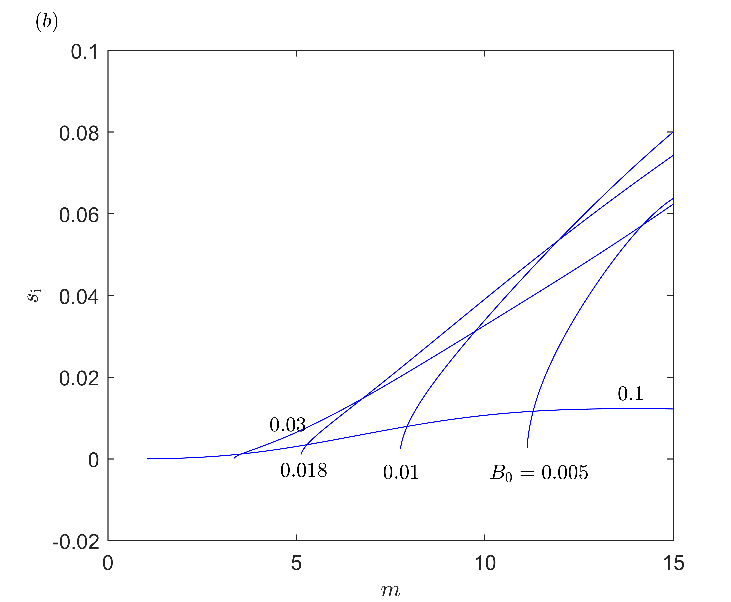}
  \caption{The dispersion relation giving $s$ for different magnetic field strengths $B_0$ at $\eta=10^{-4}$, corresponding to $\eta_\star=0.19$. The other parameters are the same as figure \ref{F5}.  The real part $s_{\mathrm{r}}$ of the growth rate is shown in panel (a) and the imaginary part  $s_{\mathrm{i}}$ in panel (b). Red lines represent real solutions,  and blue lines represent complex solutions.   }\label{F6}
\end{figure}

Figure \ref{F6} shows $s$ calculated from the dispersion relation at a moderate magnetic diffusivity $\eta=10^{-4}$, equal to the viscosity $\nu=10^{-4}$. The red lines correspond to   real solutions for $s$, and the blue lines give complex solutions $s=s_\mathrm{r}+\mathrm{i}s_\mathrm{r}$. Unfortunately,  we do not have analytical results for a comparison in this regime and only show numerical solutions. The magnetic field again has a stabilising effect, but the behaviour of the dispersion relation is significantly different from figure \ref{F5}. In particular, there are two branches of real modes which can be unstable; the lower branch originates from the stable modes at $B_0=0$, corresponding to the dissipation of mean field governed  by (\ref{3.34}). When two branches of real modes merge at a certain wavenumber (e.g. $m\approx 7$ for $B_0=0.01$),   a complex mode branches out. As the magnetic field increases, the waveband of real modes shrinks and that of complex modes broadens. Complex modes can also have $s_\mathrm{r}>0$ and thus be unstable, but the real modes generally have larger growth rates. A magnetic field reduces the growth rates of both the real and complex modes, and suppresses instability when strong enough. The imaginary part $s_\mathrm{i}$ first increases with the field, and then decreases when the field becomes strong. Our dispersion relation in this case is similar to figure 2 of \citet{Constantinou18}.

\begin{figure}
  \centering
  \includegraphics[width=0.6\linewidth]{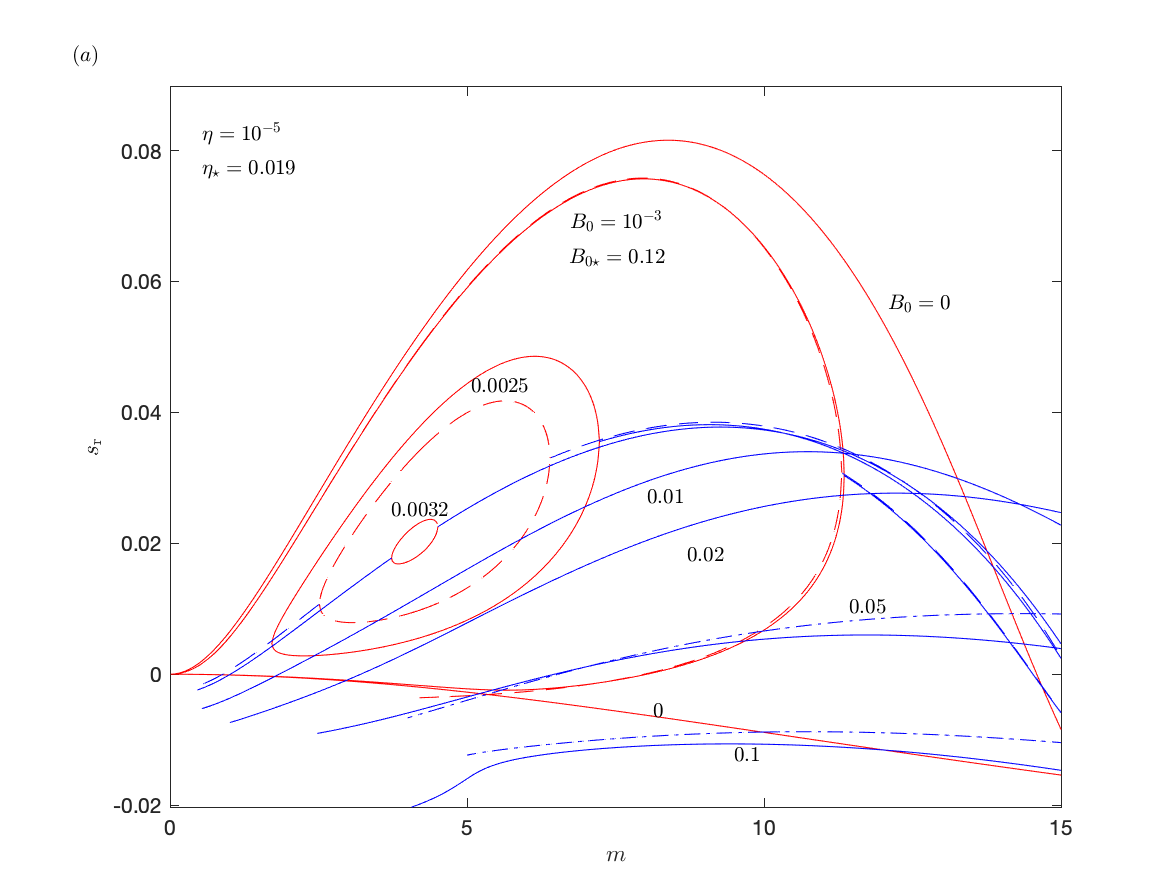}
    \includegraphics[width=0.6\linewidth]{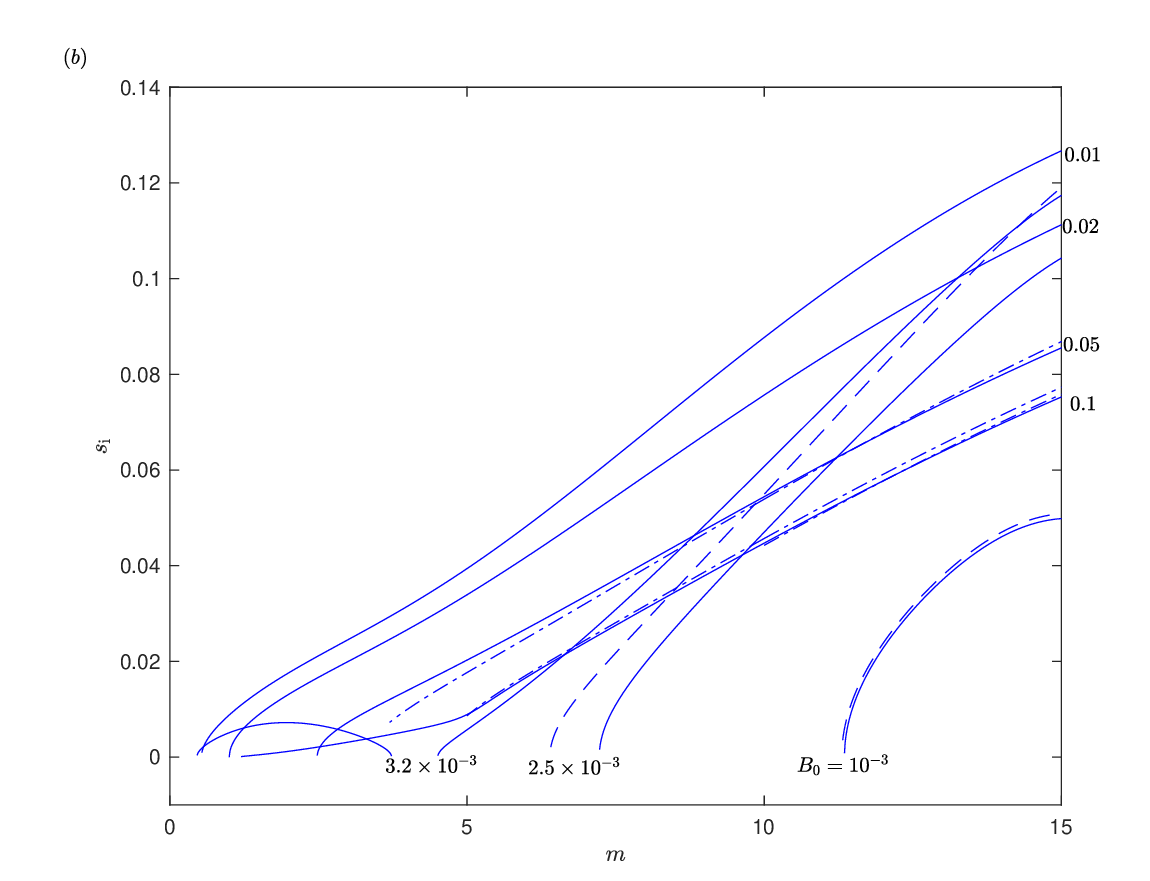}
  \caption{The dispersion relation giving $s$ for different magnetic field strengths $B_0$ at $\eta=10^{-5}$, corresponding to $\eta_\star=0.019$. The other parameters are the same as figure \ref{F5}.  The solid lines are numerical solutions; the dash and dash--dot lines are the solutions of the asymptotic dispersion relations (\ref{3.41}) for weak field and (\ref{3.44})  for strong field, respectively.  }\label{F7}
\end{figure}

Figure \ref{F7} shows the dispersion relation at a weak diffusivity $\eta=10^{-5}$, corresponding to $\eta_\star=0.019$. In this situation, our approximate solutions give good predictions for different regimes: the dashed lines are the results of the asymptotic dispersion relation (\ref{3.41}) for weak magnetic field, and the dash--dot lines are the result of (\ref{3.44}) for strong field. The dispersion relation behaves very differently from the previous two cases (figures \ref{F5} and \ref{F6}). For weak field $B_0\le 0.0032$, the curves for real modes form a family of closed loops,  and complex modes branch out at the left and right edges of each loop.  The loops shrink significantly as the field is increased from low levels, a behaviour that can be predicted from equation  (\ref{3.41}), indicating that the Maxwell stress is responsible for the stabilising effect. The loops of real modes disappear at $B_0\approx 0.0033$, and for fields stronger than this all unstable modes are complex. The growth rate of complex modes then reduces as the field becomes stronger, but less dramatically: instability is suppressed only when $B_0$ reaches the order of $0.1$, corresponding to $B_{0\star}=12$. The behaviour at such strong fields is well captured by the asymptotic dispersion relation (\ref{3.44}), indicating that the interaction between the mean flow and mean field is the main source for the instability.

\begin{figure}
  \centering
  \includegraphics[width=0.7\linewidth]{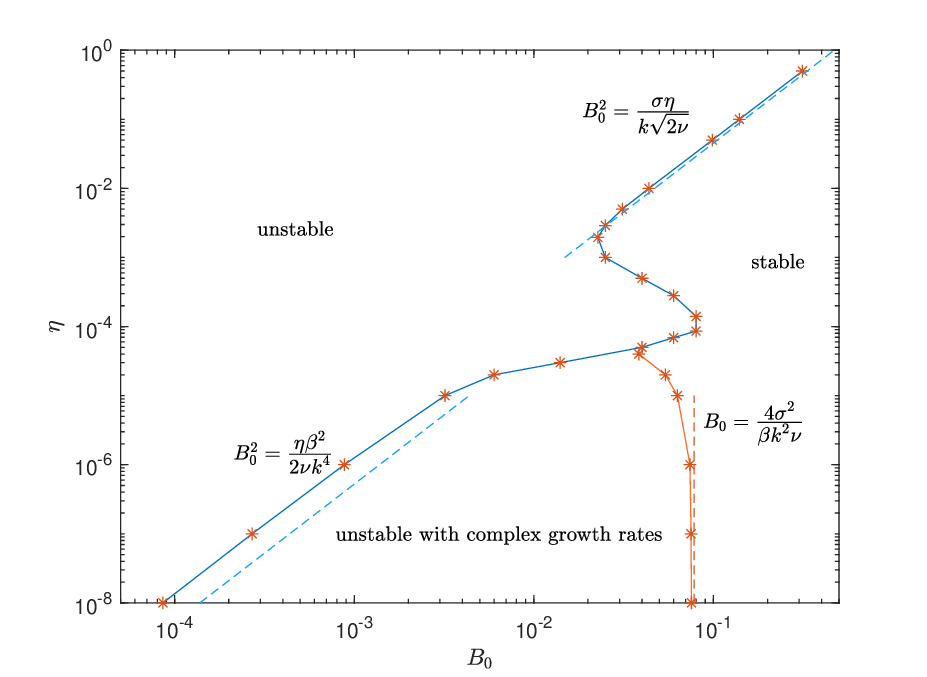}
  \caption{Stability diagram of the mean flow/field zonostrophic instability  in the $(B_0,\eta)$-plane. The  parameters are $\beta=5$, $\sigma=0.05$, $\nu=10^{-4}$, $\mu=0$, $k=16$ corresponding to $\beta_\star=2.3$, $\nu_\star=0.19$.
The region `unstable with complex growth rates' means all real growth rates are negative, but complex growth rates have positive real parts. Dashed lines are the indicated scaling laws.   
}\label{F8}
\end{figure}

In figure \ref{F8}, we show the stability diagram in the $( B_0, \eta )$ plane for zonostrophic instability in the MHD system, summarising the stable and unstable regions given in  figures \ref{F5}--\ref{F7}.  The boundaries between the stable and unstable regions are indicated by the solid lines, with data points represented by the stars. Scaling laws (to be derived shortly) are plotted in dashed lines.   If the magnetic field $B_0$ is strong enough, it can suppress the instability for any prescribed $\eta$ and the flow is stable. On the other hand, for any fixed $B_0$, if the diffusivity $\eta$ is high enough, it can counteract the effect of the field and we recover the hydrodynamic instability. There is a region labelled `unstable with complex growth rates', which corresponds to the situation in figure \ref{F7}($a$): for $0.0032<B_0<0.5$, there are no real unstable modes but complex modes may have positive growth rates. For typical flow instability problems, this region would be simply regarded as `unstable', but there are uncertainties around its interpretation for our problem which concern the statistical behaviour of zonostrophic instability. Thus we regard this region with caution and will discuss its subtle properties later on.

We now derive scaling laws for the stability boundary based on our reduced dispersion relations, which will provide new insights and general conclusions.   Again, we assume $\mu_\star\ll\nu_\star\ll1$ while $\beta_\star$ is of order of unity or larger. At high magnetic diffusivity, the instability threshold can be predicted by the asymptotic dispersion relation (\ref{3.39}). For neutral stability $s=0$, if we balance the viscosity term on the left-hand side with the right-hand side, assuming small $m_\star$ (as we see in figure \ref{F5}), we have
\begin{equation}
  \nu_\star m_\star^2\sim \frac{m_\star^2}{2\times   ({B_{0\star}^2}/{\eta_\star})\times ({B_{0\star}^2}/{\eta_\star})\times 2}\times 2,
\end{equation}
 which yields a scaling law
\begin{equation}
  B_{0\star}^2\sim \frac{\eta_\star}{\sqrt{2\nu_\star}}\qquad \mathrm{or} \qquad  B_0^2\sim \frac{\eta \sigma}{k\sqrt{2\nu}}\, . \label{3.48}
\end{equation}
This scaling law provides the very precise prediction seen in the top right of figure \ref{F8}.   

At low magnetic diffusivity, the boundary between the region of `unstable' and `unstable with complex growth rates' corresponds to when the loops of real modes disappear in figure \ref{F7}($a$), and may be estimated using  the asymptotic dispersion relation (\ref{3.41}). When the loops disappear, $s_\star$ approximately reduces by half, so we may very roughly estimate that the magnetic term in (\ref{3.41}) reaches half of the hydrodynamic term. Noting $s_\star\gg \nu_\star$ and $m_\star$ is relatively small when the loop disappears, we have the balance
\begin{equation}
  \frac{m_\star^2  B_{0\star}^2}{2\eta_\star \beta_\star^2 s_\star}\sim \frac{1}{2}\times \frac{m_\star^2}{2\nu_\star s_\star},
\end{equation}
and this provides the scaling law
\begin{equation}\label{3.50}
  B_{0\star}^2\sim \frac{\eta_\star \beta_\star^2}{2\nu_\star}\qquad \mathrm{or}\qquad B_{0}^2\sim \frac{\eta \beta^2}{2\nu k^4}.
\end{equation}
As seen in figure \ref{F8}, this scaling law also gives fairly good predictions, and  agrees qualitatively with that found by \cite{Tobias07} and \cite{Constantinou19},  as we will elaborate subsequently.  

 Finally, the boundary between the region of `unstable for complex growth rates' and `stable' corresponds to when $s_\mathrm{r}$ of the complex mode in figure \ref{F7}$(a)$ changes sign, which can be described by the reduced dispersion relation (\ref{3.44}). Due to the complicated expressions for $N_{UB\star}$ and $N_{BU\star}$, we are not able to factor out the real part of $s_\star$, and we will only roughly estimate the threshold based on the orders of terms. Numerical solutions suggest that at the stability threshold, the purely imaginary $s_\star$ is of the order of $\nu_\star$ which is small, hence we can deduce
\begin{equation}
  N_{UB\star},\ N_{BU\star}=O\left(\frac{1}{\beta_\star B_{0\star}}\right).
\end{equation}
Note that the $O(\nu^{-1}_\star)$ terms are cancelled by the c.c.e.$s$. Then balancing the left-hand and right-hand sides of (\ref{3.44}) indicates that $B_{0\star}\propto (\beta_\star\nu_\star)^{-1}$. At this point, we can only use data fitting to find the constant factor in this relation: we find the factor $4$ fits the numerical solution well. Hence we have the scaling law:
\begin{equation} \label{3.52}
  B_{0\star}\sim \frac{4}{\beta_\star\nu_\star}\, ,\qquad B_0\sim \frac{4\sigma^2}{\beta \nu k^2}\, ,
\end{equation}
which appears in figure \ref{F8} as the vertical dashed line. Although we have fitted the constant to obtain this law, the analysis reveals key underlying physics: we have $B_{0\star}\propto \nu_\star^{-1}\gg1$, showing that complex unstable modes survive strong magnetic fields, and that as $\eta_\star$ becomes small it no longer affects the stability boundary.

For moderate magnetic diffusivity  $\eta\sim 10^{-4}  = \nu$, we do not have an asymptotic dispersion relation, unfortunately. 
We see that the stability behaviour in  figure \ref{F8}  is rather complicated: the three regions are all present, and the boundaries wobble as $\eta$ increases.  The explanation may be that since the magnetic diffusivity is similar in magnitude to the viscosity, the interaction between the flow and the field is very active. All of the mechanisms that we have identified (e.g. increased viscosity, Maxwell stress, and mean flow--mean field interaction) are all present. In other words, none of the terms or effects in (\ref{3.28}) and (\ref{3.29}) may be neglected. 

\begin{figure}
  \centering
  \includegraphics[width=0.44\linewidth]{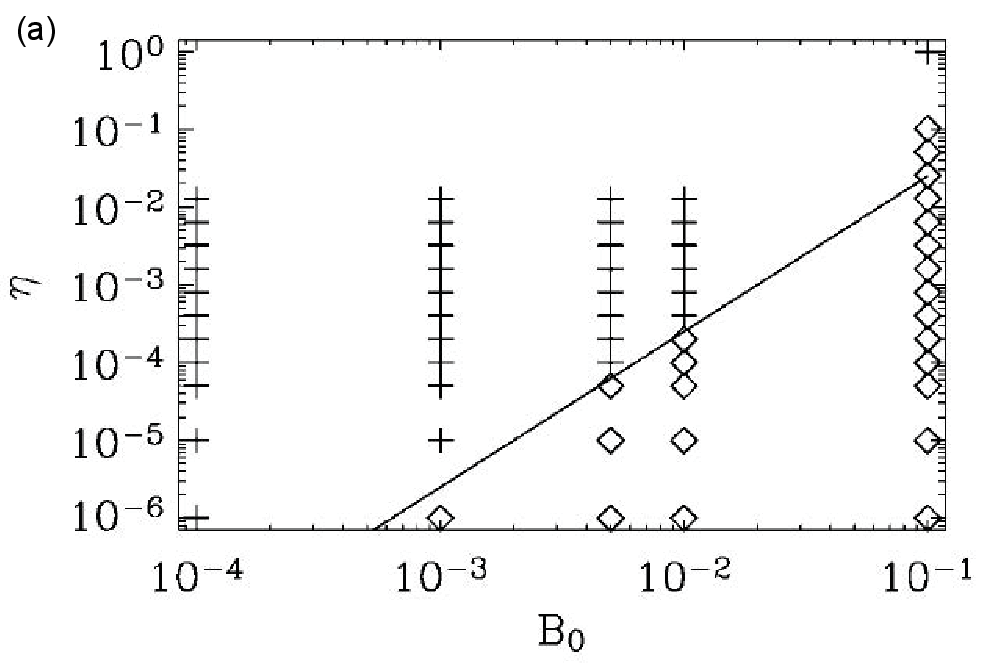}
   \includegraphics[width=0.3\linewidth]{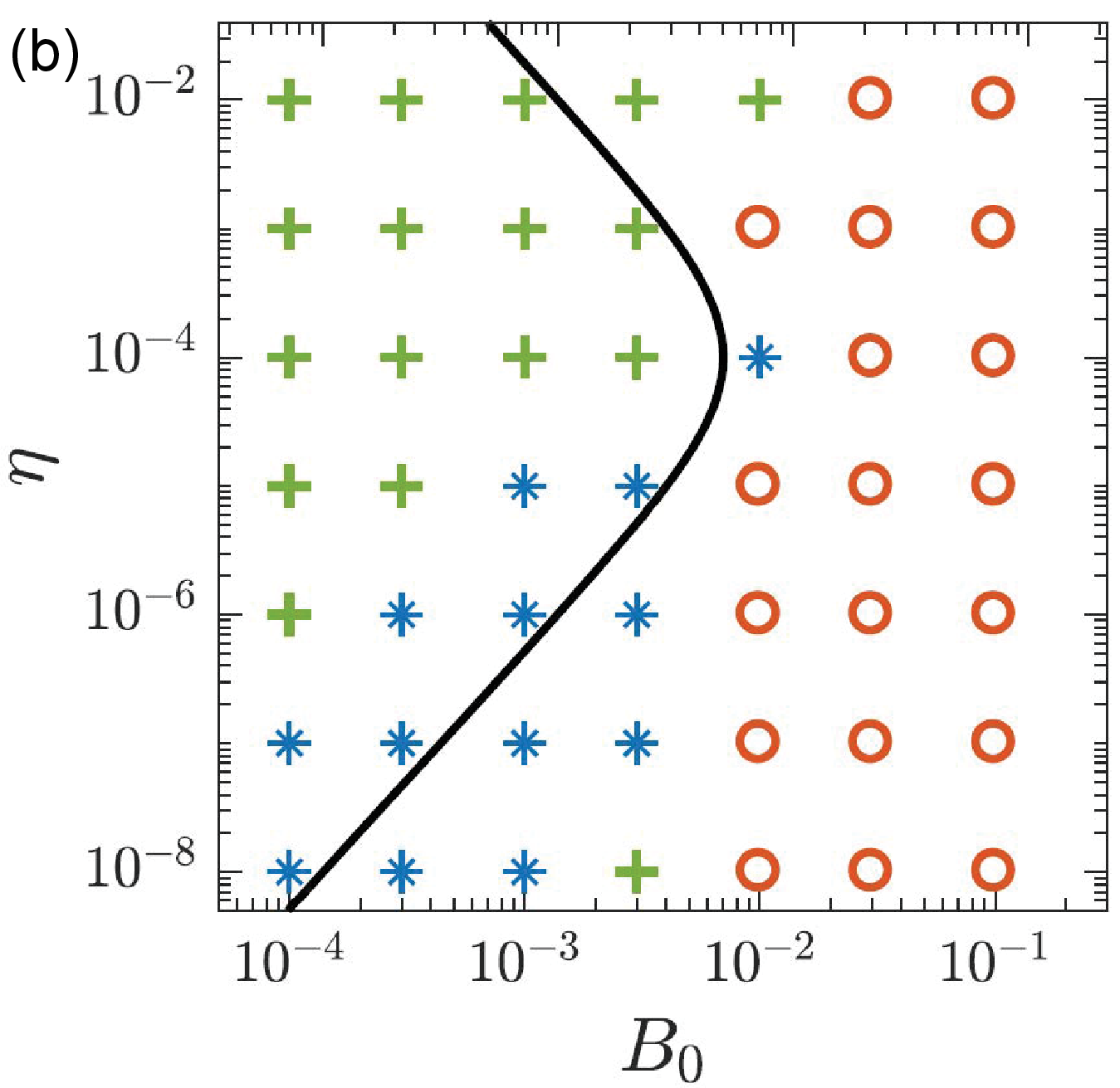}
  \caption{Stability diagram in the $(B_0, \eta)$ plane taken from ($a$) \citet{Tobias07} and $(b)$ \citet{Constantinou18} (figure reproduced with permission from the authors). In figure ($a$), the plus signs represent simulations in which large-scale zonal flows emerge, and the diamond symbols simulations in which they  do not. The solid line is defined by $B_0^2/\eta=\mathrm{constant}$. In figure ($b$), the plus signs represent unstable modes with real growth rate, the stars represent complex growth rates, and the circles represent stable modes. The solid line is defined by
 (\ref{eqCPboundary}). 
  }\label{F9}
\end{figure}

\citet{Tobias07} and \citet{Constantinou18} also considered the stability diagram in the $(B_0,\eta)$ parameter plane. We show their results in figure \ref{F9} and compare them to our figure \ref{F8}. \citet{Tobias07} performed numerical simulations and examined the conditions for which the large-scale zonal flows emerge. Their results are shown in figure \ref{F9}($a$), where a plus or diamond sign represents conditions for which a large-scale zonal flow did or did not emerge, respectively. From these data points, they found that the boundary between these two situations obeys a scaling law $B_0^2/\eta=\mathrm{constant}$. Compared to our study, this scaling law corresponds to our  result (\ref{3.50}) at low magnetic diffusivity, which also obeys $B_0^2\propto \eta$. 

\cites{Constantinou18} study was based on a zonostrophic instability analysis. One of their results is shown in figure \ref{F9}($b$), where a plus sign represents a real and positive growth rate, a star represents a complex growth rate, and a circle represents a real negative growth rate. We see the structure of the three regions is similar to our figure \ref{F8}  around $\eta\sim 10^{-4}$ (however, we note a difference in definition: their region of complex growth rates counts those with both positive and negative real part, while ours only includes those with positive real part). Their empirical boundary is defined by
\begin{equation}\label{eqCPboundary}
  \frac{\omega_A^2}{\omega_R^2}\frac{1+Pr_m^2}{Pr_m}=1,
\end{equation}
where $\omega_A$ and $\omega_R$ are the frequencies of shear Alfv\'{e}n and undamped Rossby waves, respectively, and $Pr_m=\nu/\eta$ is the magnetic Prandtl number. Translated to our notation, this scaling law becomes
\begin{equation}
  B_0^2=\frac{\nu\eta \beta^2}{(\eta+\nu)^2k^4}\,.
\end{equation}
For small magnetic field strengths, it agrees with our scaling law (\ref{3.50}) up to a constant factor. \cite{Constantinou18} inferred this scaling law from the form of the spatio-temporal correlation function of the magnetic field.  Our derivation further clarifies the physics.
 
 We will return to the discussion of the region of `unstable with complex growth rates' in our figure \ref{F8} shortly. \citet{Constantinou18} also reported complex growth rates with a positive real part. However, compared to figure \ref{F9}($a$), this region seems to correspond to the conditions where \citet{Tobias07} found no zonal flow forms, i.e. which should be regarded as zonostrophically stable.  The properties of these modes therefore remain curious. We will investigate the  behaviour of these complex modes using numerical simulations in the next section, considering particularly the stochastic behaviour of the flow and field.

\subsection{Numerical simulation} \label{S3.6}

We now perform numerical simulations for the MHD flow governed by equations (\ref{3.1}) and (\ref{3.2}). We use a pseudospectral method, as described for the hydrodynamic case, except that for weak diffusivity $\eta=10^{-5}$ corresponding to $\eta_\star=0.019$, we use a higher resolution of $512\times 512$ mesh points. We use the same initial condition $\psi=10^{-4}\cos (my)$ for the flow, and there is no magnetic perturbation, $B=0$ initially; only the uniform background field $B_0$ is present. We will consider various values of $B_0$ and $\eta$, as studied in figures \ref{F5}--\ref{F7}, and use the same values for other  parameters, namely $k=16$, $m=5$, $\sigma=0.05$, $\beta=5$, $\mu=0$, $\nu=10^{-4}$, corresponding to $\beta_\star=2.3$, $\nu_\star=0.19$, $m_\star=0.31$, as we did in the purely hydrodynamic simulation in \S 2.5.

\begin{figure}
  \centering
  \includegraphics[width=1\linewidth]{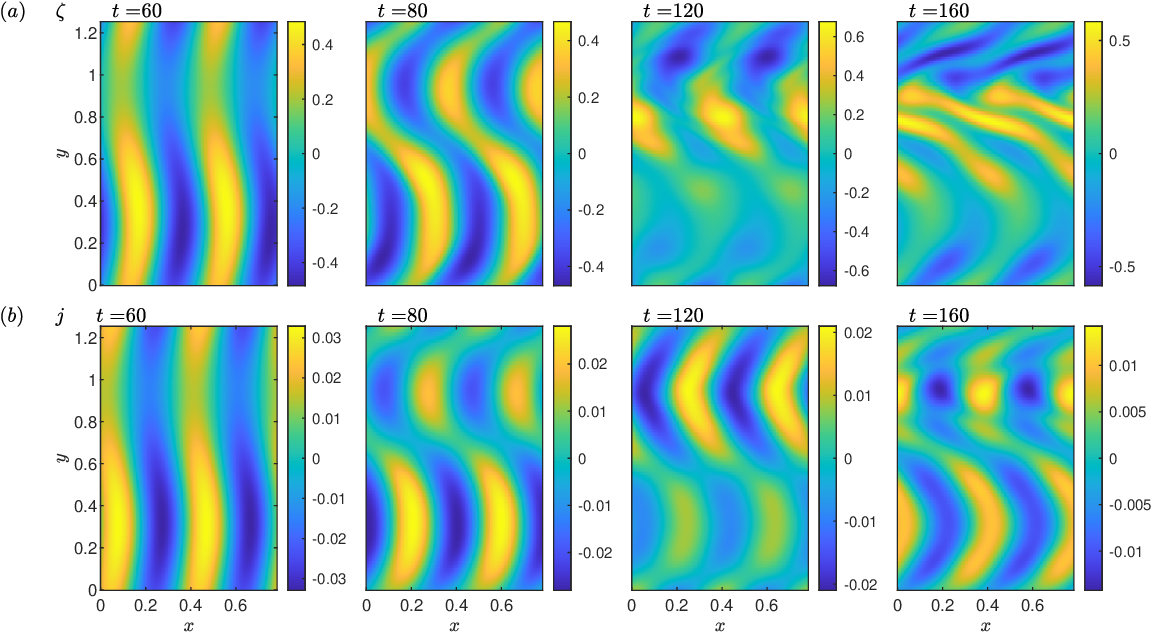}
  \includegraphics[width=0.535\linewidth]{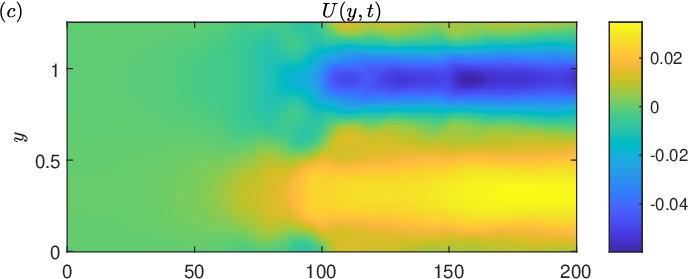}
  \includegraphics[width=0.455\linewidth]{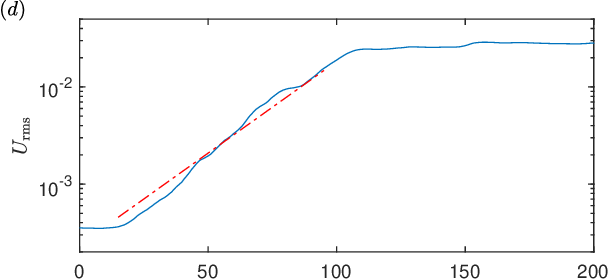}
    \includegraphics[width=0.535\linewidth]{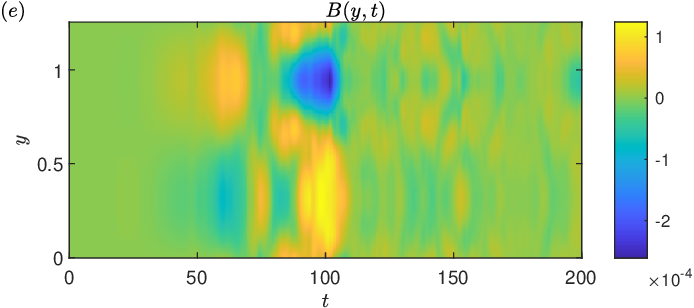}
  \includegraphics[width=0.455\linewidth]{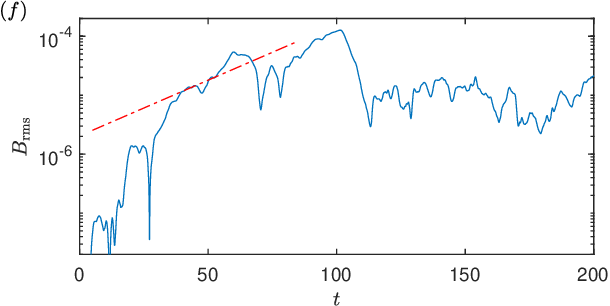}
  \caption{Numerical simulation for the MHD flow for $B_0=0.01$, $\eta=10^{-2}$, $\beta=5$,  $k=16$, $\nu=10^{-4}$, $\mu=0$, $\sigma=0.05$, $m=5$, corresponding to $\beta_\star=2.3$, $\nu_\star=0.19$, $B_{0\star}=1.2$, $\eta_\star=19$, $m_\star=0.31$. Panels $(a)$ and $(b)$ show the evolution of vorticity $\zeta$ and current density $j$; panels $(c)$ and ($e$) are the Hovm\"{o}ller diagram for the mean flow $U(y,t)$ and field $B(y,t)$, and panels $(d)$ and $(f)$ show the rms value of $U$ and $B$, respectively; solid lines are the results of the numerical simulation, and the straight dashed-dot lines are the predictions of the zonostrophic instability  with growth rate $s=4.35\times 10^{-2}$. 
      }\label{M1}
\end{figure}

We first study the case of relatively high magnetic diffusivity $\eta=10^{-2}$. The evolution of the vorticity $\zeta$ and  current density $j$ are shown in figure \ref{M1}($a$, $b$), and the evolution of the mean flow $U$ and mean field $B$ are shown in figure \ref{M1}($c$--$f$). The behaviour of $\zeta$ and $U$ is similar to the hydrodynamic case:  zonostrophic instability causes exponential growth of the zonal flow, forming two zonal jets in opposite directions. The zonal jets shear the vorticity, causing sinuous winding which becomes stronger over time. The exponential growth of $U_\mathrm{rms}$ in panel (d) agrees very well with the expectation predicted by the zonostrophic theory. The spatial structure of $j$ in panel (b) behaves in a similar way to that of $\zeta$ in panel (a), but has a much weaker amplitude. The mean field $B$ grows exponentially during the exponential growth of the zonal flow in panel (f), but then falls back to very small values. It thus appears that the field is largely controlled by the flow.

\begin{figure}
  \centering
  \includegraphics[width=0.45\linewidth]{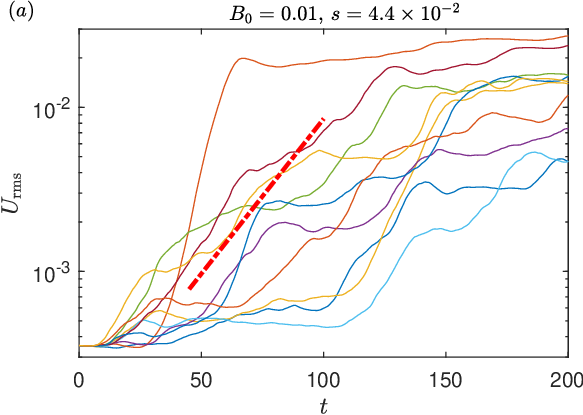}
   \includegraphics[width=0.45\linewidth]{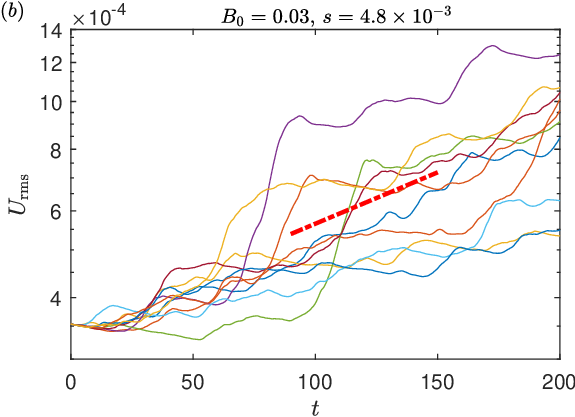}
 \includegraphics[width=0.45\linewidth]{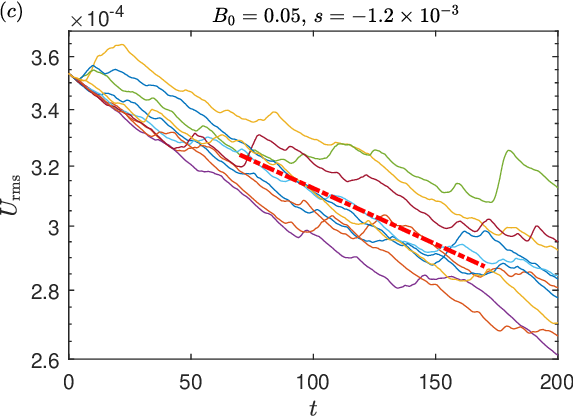}
  \includegraphics[width=0.45\linewidth]{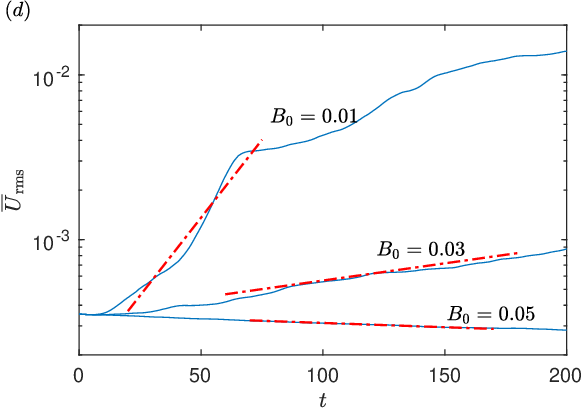}
  \caption{Temporal evolution of $U_\mathrm{rms}$ for $\eta=10^{-2}$ ($\eta_\star=19$) and ($a$) $B_0=0.01$ $(B_{0\star}=1.2)$, $(b)$ $B_0=0.03$, $(c)$ $B_0=0.05$. The other parameters are the same as in figure \ref{M1}. Solid lines represent 10 realisations of the numerical simulation of the stochastic flow,  and the thick dash--dot lines represent the prediction of the zonostrophic instability. In panel $(d)$, the ensemble average of the ten realisations for each $B_0$ is shown by solid lines, compared to the straight dash--dot line representing the prediction of the zonostrophic instability.   
  }\label{UM1}
\end{figure}

To further explore the stochastic behaviour of the MHD system, we consider three different magnetic field strengths $B_0=0.01, 0.03, 0.05$ at the same $\eta=10^{-2}$, and  run ten simulations for each case;  see figure \ref{F5} for growth rates obtained from the dispersion relation in this case. The numerical results for $U_\mathrm{rms}$ are shown in figures \ref{UM1}$(a$--$c)$. We also plot a thick dash--dot line to indicate the exponential growth of the expectation, as predicted by the zonostrophic instability theory. The value of  the growth rate $s$ is indicated in the title of each plot. The variability in behaviour and growth across the ten different realisations is striking and so in figure \ref{UM1}$(d)$, we show the ensemble averages in solid lines to compare with the zonostrophic instability theoretical prediction in dash--dot lines. There is good agreement for all  three cases, confirming the theory. The stabilising effect of the magnetic field is  also clearly demonstrated. In figures \ref{UM1}$(a$--$c)$, although there is significant variability, the theoretical prediction for the expectation generally agrees with the growth or decay in each realisation, thus providing support for the full ergodic assumption (\ref{ergo2}).

\begin{figure}
  \centering
  \includegraphics[width=1\linewidth]{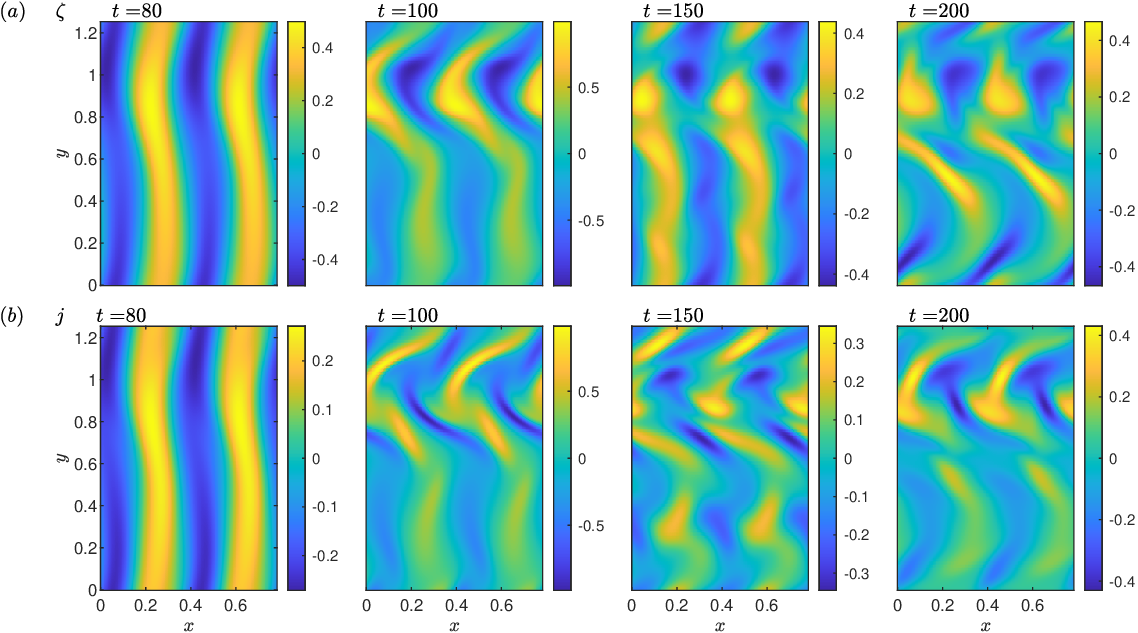}
  \includegraphics[width=0.535\linewidth]{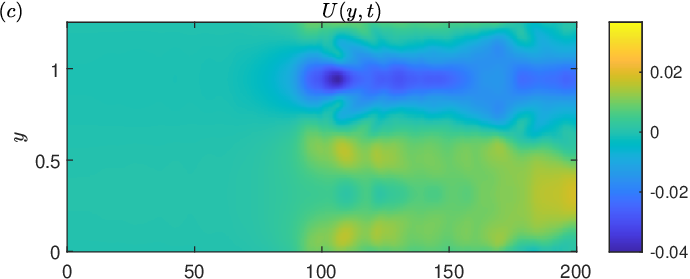}
  \includegraphics[width=0.455\linewidth]{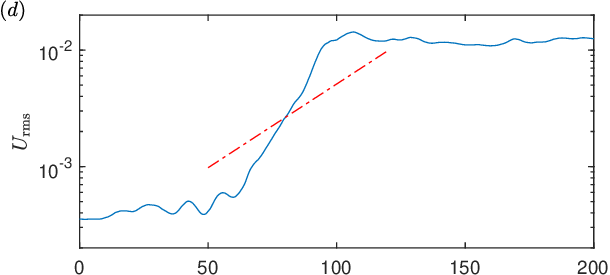}
    \includegraphics[width=0.535\linewidth]{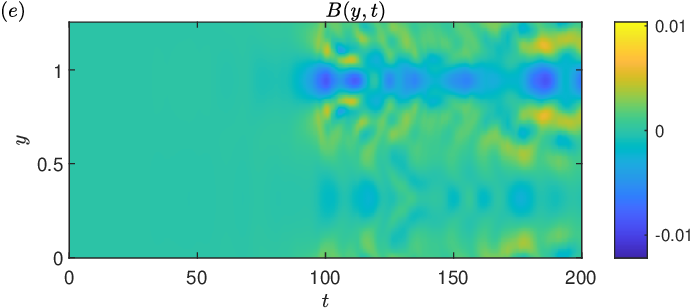}
  \includegraphics[width=0.455\linewidth]{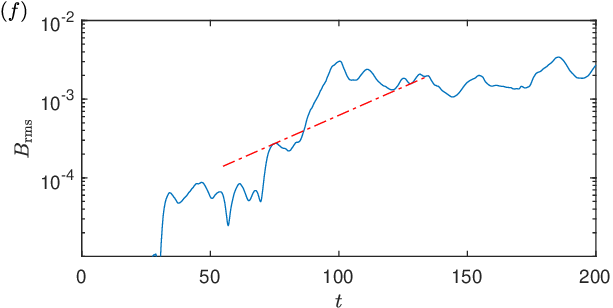}
  \caption{Numerical simulation for the MHD flow for $B_0=0.01$, $\eta=10^{-4}$ corresponding to $B_{0\star}=1.2$, $\eta_\star=0.19$, with the theoretical growth rate $s=3.31\times 10^{-2}$.  The other parameters and the contents of the panels are the same as in figure \ref{M1}.    }\label{M2}
\end{figure}

Next, we  consider the zonostrophic instability at a moderate diffusivity $\eta=10^{-4}$; results from the dispersion relation for this case have been set out in figure \ref{F6}. In figure \ref{M2}, we present the evolution of $j$, $\eta$, $U$ and $B$ for one realisation with $B_0=0.01$. Zonal flows again emerge as a result of zonostrophic instability, but the weaker magnetic diffusivity results in a stronger influence of the magnetic field. For example, the strengths of $j$ and $B$ are now of the same order of $\zeta$ and $U$, respectively, indicating similar importance of the flow and the field. The spatial pattern of $j$ in panel (b) is characterised by elongated thin filaments, somewhat different from the vortices seen for $\zeta$ in panel (a). The growth of $U_\mathrm{rms}$ in panel (d) is somewhat faster than the expectation that the zonostrophic instability predicts, and the  agreement is now only qualitative.


\begin{figure}
  \centering
  \includegraphics[width=0.45\linewidth]{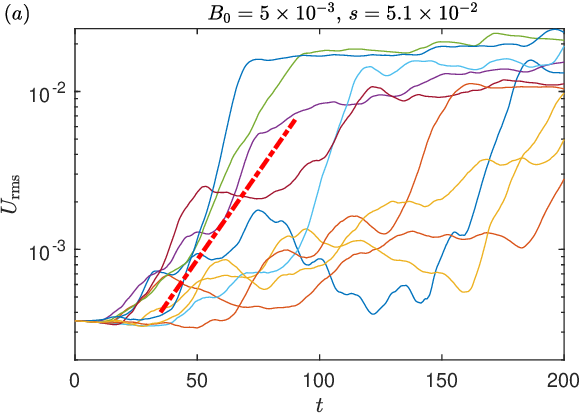}
   \includegraphics[width=0.45\linewidth]{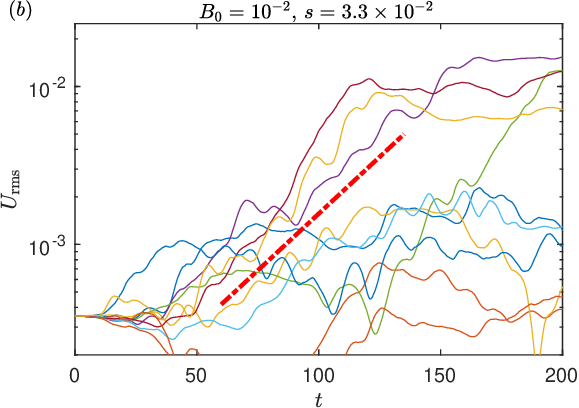}
 \includegraphics[width=0.45\linewidth]{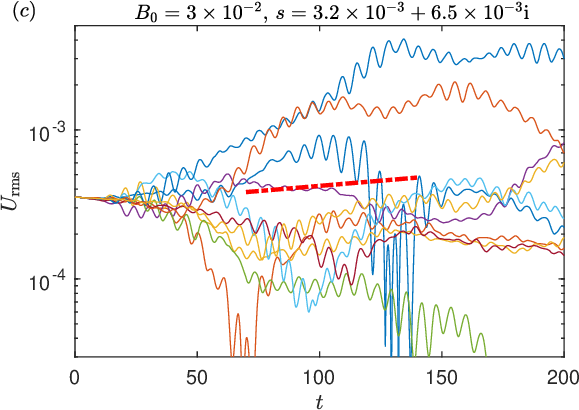}
  \includegraphics[width=0.45\linewidth]{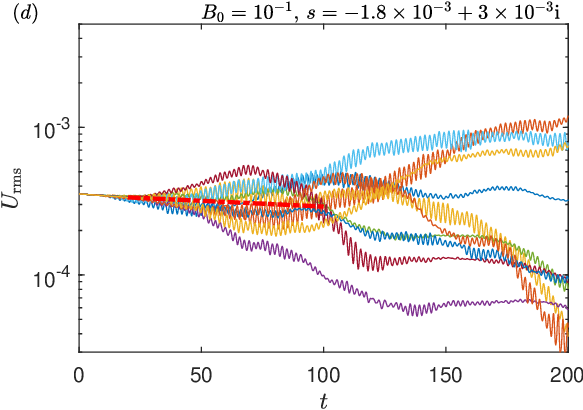}
  \includegraphics[width=0.45\linewidth]{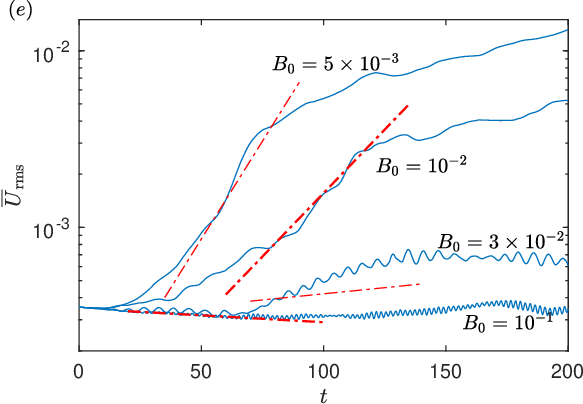}
  \caption{Temporal evolution of $U_\mathrm{rms}$ at $\eta=10^{-4}$ ($\eta_\star=0.19$) and ($a$) $B_0=5\times 10^{-3}$, $(b)$ $B_0=10^{-2}$ ($B_{0\star}=1.2$), $(c)$ $B_0=3\times 10^{-2}$, $(d)$ $B_0=10^{-1}$. The other parameters are the same as in figure \ref{M1}. Solid lines represent 10 realisations of the numerical simulation of the stochastic flow,  and the thick straight dash--dot lines represent the prediction of the zonostrophic instability. In panel $(e)$, the ensemble average of the ten realisations for each $B_0$ is shown by solid lines, compared to the straight dash--dot line representing the prediction of the zonostrophic instability.  }\label{UM2}
\end{figure}

Following the typical realisation  in figure \ref{M2} for a moderate diffusivity $\eta=10^{-4}$ with $B=0.01$, we have undertaken a series of runs. Figure \ref{UM2} shows ten realisations for each of the four magnetic field strengths $B_0=5\times 10^{-3}, 10^{-2}, 3\times 10^{-2}$ and $10^{-1}$ in panels $(a$--$d)$; the theoretical growth rate of the expectation is plotted using dash--dot lines. The ensemble average for each field strength is shown in panel $(e)$ and also compared with theory. The growth rates $s$ for $B_0=3\times 10^{-2}$ and $10^{-1}$ in panels $(c, d)$ are complex, and we only plot the exponential growth or decay from the real part of $s$; we will elaborate more on this issue shortly. For very weak magnetic field $B_0=5\times 10^{-3}$ in panel $(a)$, each realisation has a growth rate reasonably close to that the theory predicts for the expectation, with good agreement for the ensemble average in panel $(e)$,  similar to the hydrodynamic case.  As $B_0$ increases to $10^{-2}$ in panel $(b)$, for realisations where exponential growth is prominent, the theoretical result still captures the behaviour fairly well, resulting in good agreement again for the average in panel $(e)$.  However, there are also many realisations in panel $(b)$ where the zonal flow does not grow (up to the time $t=200$ that we simulate), and the system is now showing an increased degree of randomness.

At a stronger background field $B_0=3\times 10^{-2}$, where $s=3.2\times 10^{-3}+6.5\times 10^{-3}\mathrm{i}$ becomes complex, individual realisations become more chaotic, in panel $(c)$: the growing realisations often have much larger growth rates than the theory for the expectation, while the decaying ones may reach very small amplitudes. Their ensemble average in  panel $(e)$ also shows a poorer agreement with the expectation. Finally, at the largest  $B_0=0.1$, the theory predicts that the expectation should decay; individual realisations may again behave differently, but the chaotic nature seems to have been weakened. At earlier times, the ensemble average has good agreement with the expectation, whilst at later times, occasional growth of some realisations make the ensemble average diverge from the expectation. 

In summary, at this lower value $\eta=10^{-4}$ of the magnetic diffusivity, zonal flows of individual realisations have a higher degree of randomness and their growth may diverge from the expectation predicted by the theory. The full ergodic assumption for the mean flow and field stated in (\ref{ergo2}) is therefore very questionable here, and in fact does not appear to operate in any meaningful, qualitative, way. However our instability analysis does not use the full ergodic assumption in the form (\ref{ergo2}). Rather, we used the partial ergodic assumption (\ref{assumption2}) which allows variation of individual realisations from the expectation, and is thus a better approximation in this situation. The use of the partial ergodic assumption is supported by the  fairly good agreement between  the expectation from the theory and the ensemble average of the simulations.


In figures \ref{UM2}$(c, d)$, we also observe prominent high-frequency oscillations of the zonal flow. We find their frequencies are very well described by
\begin{equation}\label{O}
  \omega=2k B_0,
\end{equation}
i.e., the zonal flow oscillates at twice the characteristic Alfv\'{e}n wave frequency. There is also a remnant of such an oscillation in figure \ref{UM2}$(b)$, though the agreement with (\ref{O}) is not as good since the oscillations are less distinguishable. Note that the frequency of these oscillations is not  related to the imaginary parts of $\lambda_{1\pm}$ and $\lambda_{2\pm}$ (cf. (\ref{3.13}) and (\ref{3.27})) or $s$.

\begin{figure}
  \centering
  \includegraphics[width=1\linewidth]{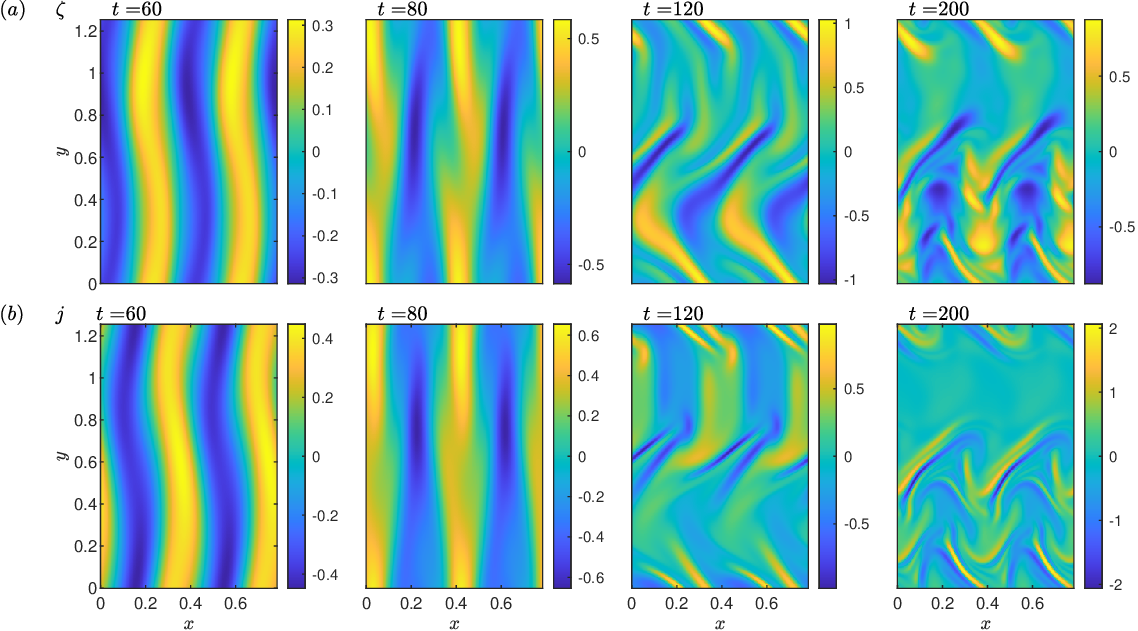}
  \includegraphics[width=0.54\linewidth]{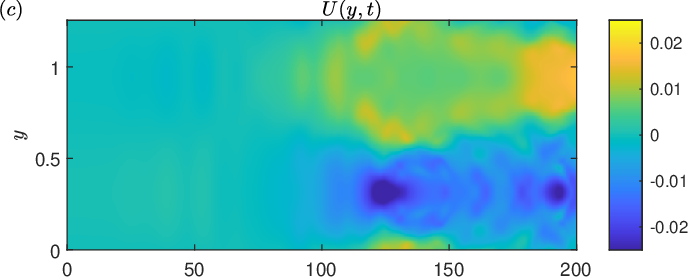}
  \includegraphics[width=0.452\linewidth]{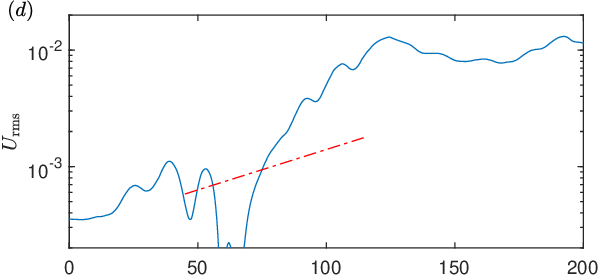}
    \includegraphics[width=0.54\linewidth]{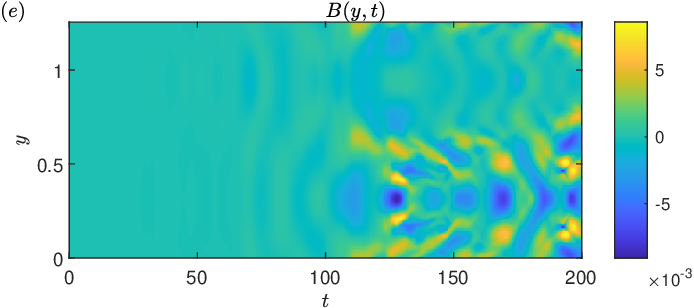}
  \includegraphics[width=0.452\linewidth]{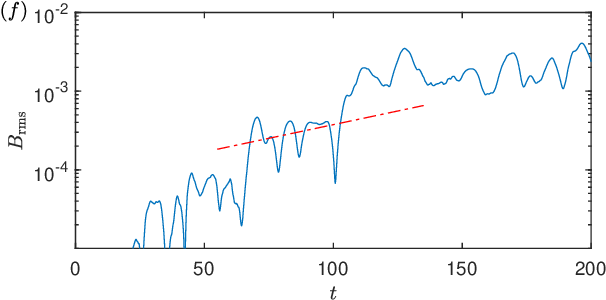}
  \caption{Numerical simulation for the MHD flow for $B_0=0.01$, $\eta=10^{-5}$, corresponding to $B_{0\star}=1.2$, $\eta_\star=0.019$, with theoretical growth rate $s=1.60\times 10^{-2}+3.94\times 10^{-2}\mathrm{i}$.  The other parameters and panels are the same as in figure \ref{M1}. In panels $(d)$ and $(f)$, only the real part of $s$ has been used to plot the theoretical growth of zonostrophic instability.  }\label{M3}
\end{figure}

Finally, we explore the case of very weak magnetic diffusivity $\eta=10^{-5}$, corresponding to the zonostrophic instability studied in figure \ref{F7}. In figure \ref{M3}, we show a realisation for $B_0=0.01$. The weak magnetic diffusivity renders very fine filaments in the spatial pattern of the current $j$, which also influences the pattern of $\zeta$. The exponential growth of $U_\mathrm{rms}$ and $B_\mathrm{rms}$ is much faster than that predicted by the theory for the expectation, indicating a high degree of stochasticity. The oscillations of the  mean flow and the mean field, described by (\ref{O}), become more prominent. A key point we emphasise is that as the theoretical growth rate of zonostrophic instability is $s=0.016+0.039\mathrm{i}$, this case falls into the regime of `unstable with complex growth rates' in figure \ref{F8}. Hence, in constrast to \citet{Tobias07} who attributed this region to be stable, we have a concrete example of zonostrophic instability taking place and generating zonal flows.  The data of the white noise for this realisation has been documented and is available online (see the Data Access Statement at the end of the paper).

\begin{figure}
  \centering
  \includegraphics[width=0.45\linewidth]{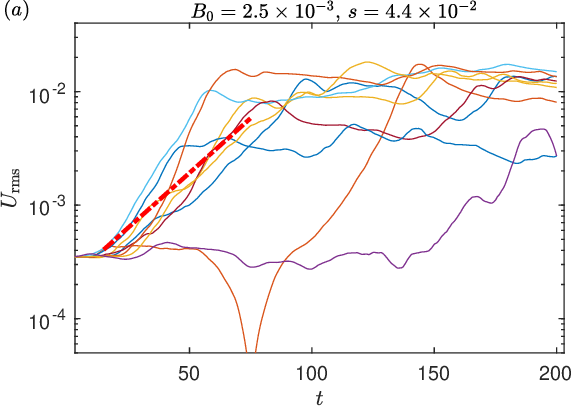}
   \includegraphics[width=0.45\linewidth]{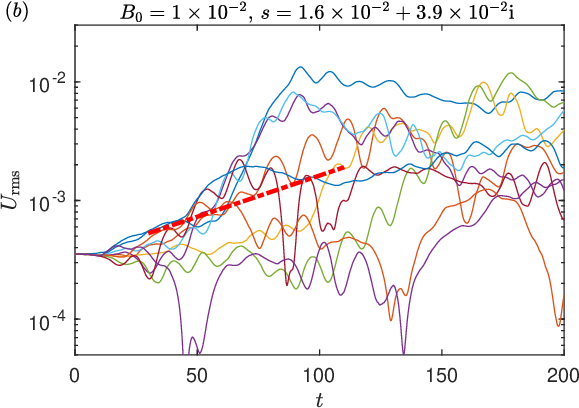}
 \includegraphics[width=0.45\linewidth]{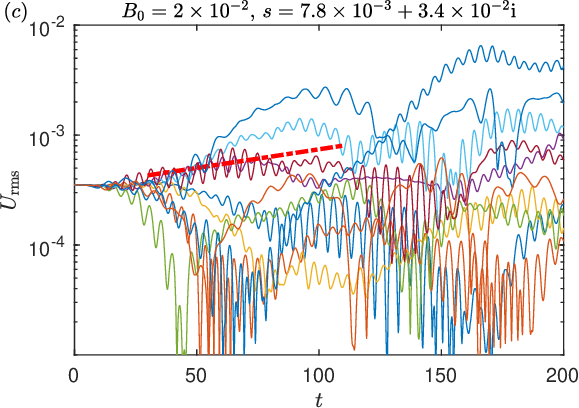}
  \includegraphics[width=0.45\linewidth]{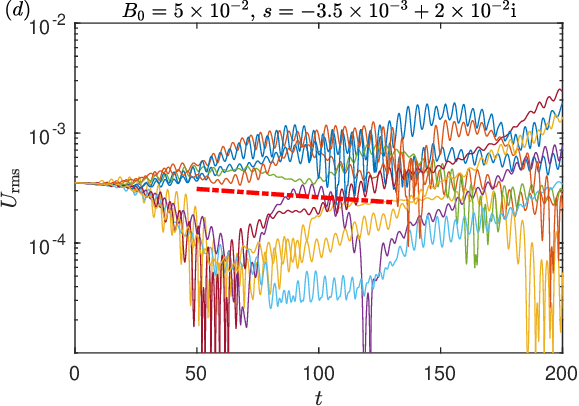}
  \includegraphics[width=0.45\linewidth]{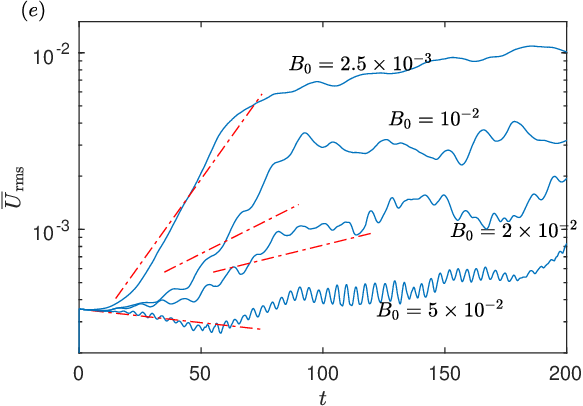}
  \caption{Temporal evolution of $U_\mathrm{rms}$ for $\eta=10^{-5}$ ($\eta_\star=0.019$) and ($a$) $B_0=2.5\times 10^{-3}$, $(b)$ $B_0=10^{-2}$ ($B_{0\star}=1.2$), $(c)$ $B_0=2\times 10^{-2}$, $(d)$ $B_0=5\times 10^{-2}$. The meanings of the plots are the same as in figure \ref{UM2}.}\label{UM3}
\end{figure}

We then show ten realisations at various magnetic field strengths $B_0=2.5\times 10^{-3}, 10^{-2}, 2\times 10^{-2}$ and $5\times 10^{-2}$ with $\eta=10^{-5}$ fixed in figure \ref{UM3}. The behaviour shows many similarities to figure \ref{UM2}, but  our main objective is to confirm the statistical behaviour of modes which are `unstable with complex growth rates' in figure \ref{F8}. At the low diffusivity of $\eta=10^{-5}$, a wide range of field strengths $B_0$ fall into this category, and as shown in figure \ref{F7}, unstable complex modes are dominant  for stronger magnetic field. In figure \ref{UM3}, the cases of $B_0=10^{-2}$ and $2\times 10^{-2}$ in panels $(b)$ and $(c)$ are unstable with complex growth rates.   As we see, the realisations for  $B_0=10^{-2}$ and $2\times 10^{-2}$ are highly chaotic. The theory developed predicts an exponential growth of the expectation, but individual realisations either have much faster growth or decay significantly; very few evolve as  theory predicts. The high frequency oscillations described by (\ref{O}) also become very strong, making the evolution even more disorganised. There appear to be more decaying realisations for $B_0=2\times 10^{-2}$ than for $B_0=10^{-2}$,  an indication of the stabilising effect of the field. The agreement between the  ensemble average over the ten realisations simulated and the growth rate of the expectation shown in panel $(e)$ 
for $B_0=10^{-2}$ and $2\times 10^{-2}$ is adequate, 
but not as good as the other two situations, i.e.\ $B_0=2.5\times 10^{-3}$ and $B_0=5\times 10^{-2}$ where the growth rates are real.  Hence we have confirmed that the modes that are unstable with complex growth rates are highly stochastic: for an individual realisation,  zonal flow may or may not emerge, and the growth rate predicted for the expectation of the mean fields has little relevance.  While the full ergodic assumption (\ref{ergo2}) clearly does not work, the partial ergodic assumption (\ref{assumption2}) is also put into doubt given the expectation from the theory does not agree so well with the ensemble average from the simulations.

Our conclusion  seems different to that of \citet{Tobias07} who suggest that the region of `unstable with complex growth rates' in figure \ref{F8} should have no zonal flow formation. We think the reason for this disparity may lie in the forcing: our simulations indicate that different stochastic forcings yield different evolution of the mean flow, so that any single simulation might not be representative. It is also possible that the spatial structure of the forcing could make a difference: our forcing only has one wavenumber in the $x$-direction, while \citet{Tobias07} used a range of wavenumbers in both the $x$ and $y$-directions. We leave the issue of more realistic forcing for further investigation.  

Despite our efforts in investigating the modes with complex growth rates, there are still many aspects that we do not fully understand, in particular the physical meaning of the imaginary part of $s$.  A straightforward interpretation is that it represents oscillation of the expectation, but in the stochastic setting all phases are possible,  presumably leading to significant cancellation. Indeed, the ensemble mean from numerical simulations does not pass through zero as the theory would otherwise predict. Mainly for this reason, we have not included the imaginary part of $s$ in comparisons with the numerical simulations. On the other hand, when $s$ is complex with a positive real part, the agreement between the theory and the simulations is not so good,  suggesting that the imaginary part of $s$ has a role that we do not yet understand.  Linked with this is the observation that when $s$ is complex, runs show highly disorganised evolution of mean flow and field; we do not know if these are related, nor can we currently explore this in depth given the expense of computing a sufficiently large ensemble. Since complex growth rates can also arise for hydrodynamic zonostrophic instability with a more complicated structure of forcing (Ruiz et al. 2016), we suspect similar phenomena could take place in these non-magnetic systems. Studying this further, perhaps via a Fokker--Planck equation, is a topic for future research. 

\section{Conclusions and remarks}

In this paper, we have studied zonostrophic instability focusing on its statistical properties and the effect of a magnetic field. We apply a stochastic forcing with its amplitude varying in the form of a white noise, which generates random waves.  Weak zonal flows  can then grow exponentially to generate strong and stable zonal jets.  We study the expectation of the zonal flow and we have derived the dispersion relation for its exponential growth.  
We have also undertaken numerical simulations of many realisations of stochastic flows to compare with theory, and examined the validity of the widely used mean-flow ergodic assumption, which assumes the zonal mean flow remains the same in different realisations. 

In the zonostrophic instability analysis, we have developed a method that does not depend on the  spatio-temporal correlation functions employed in many previous studies. Instead, we analyse stochastic waves directly taking advantage of the temporal delta-correlation property of white noise. This allows us to derive simplified dispersion relations in the limits of weak and strong magnetic field and magnetic diffusivity.  Our analysis has revealed the key role played by the temporal correlation of the stochastic waves in the zonostrophic instability. Regarding the mean-flow ergodicity, our derivation depends on a weaker assumption compared to previous studies. We assume that the mean flow can still be stochastic, but its stochasticity is uncorrelated to the stochasticity of the waves. We refer to our assumption as a `partial ergodic assumption', in contrast to the `full  ergodic assumption' where the mean flow has no stochasticity.

  The dispersion relations that we derive provide straightforward insights into the effect of the magnetic field and scaling laws for the thresholds of instability. We find that when the magnetic diffusivity is strong, the effect of the magnetic field is equivalent to increasing the viscosity of the flow. When the magnetic diffusivity is very weak, for weak uniform magnetic field, it acts through the Maxwell stress to counteract the Reynolds stress, whereas at strong magnetic field, the interaction between the mean flow and mean field is the dominant dynamics. The magnetic field mainly plays a stabilising role and inhibits zonal flows, but in the regime of weak diffusivity and strong field, the interaction between the mean flow and mean field can have a destabilising effect.  
  
In the numerical simulations, we have seen  the unstable formation of zonal flow. In order to take into account the stochastic nature of the system, we have run multiple simulations in each parameter regime. Comparing the ensemble average of the mean flow to the growth of expectation predicted by the zonostrophic theory,  we find good agreement in general. When comparing the theoretical expectation to individual realisations, we find that there is good agreement in the purely hydrodynamic case, if the magnetic field is weak or if the magnetic diffusivity is strong. Otherwise, significant differences between individual realisations and the theoretical growth of the expectation typically take place.  Specifically, when the growth rate of the zonostrophic instability is complex with positive real part, the individual realisations can have very chaotic behaviours, bearing rather weak relation to the expectation, and whether the zonal flow will emerge for any one realisation is highly unpredictable. Use of the full  ergodic assumption,  i.e. assuming the mean flow to be the same in different realisations, thus appears to be very questionable in these situations. The partial ergodic assumption, on which our analysis is based, is found to be a better approximation as it allows  stochasticity of the mean flow.  

The increased randomness of the zonal flow caused by the magnetic field  for weak magnetic diffusion deserves further research. For example, what is the mechanism by which the magnetic field makes the zonal flow more disorganised? Is there a way to quantify and analyse this, for example through the covariance of the stochastic process? We may also explore the effect of more realistic forcing, for example, an isotropic ring forcing with wave vectors in all directions. The ergodic assumption of the mean flow is very widely used in statistical studies of turbulence based on cumulant 
expansions \citep{Tobias11,Marston08}. Since we have shown that it does not always hold well for MHD,  the impact of non-ergodicity deserves further study.

\section*{Acknowledgments}

This work is supported by the EPSRC (grant EP/T023139/1), which is gratefully acknowledged. Some computational studies used the DiRAC Data Intensive service at Leicester, operated by the University of Leicester IT Services, which forms part of the STFC DiRAC HPC Facility ({\tt www.dirac.ac.uk}). The equipment was funded by BEIS capital funding via STFC capital grants ST/K000373/1 and ST/R002363/1 and STFC DiRAC Operations grant ST/R001014/1. DiRAC is part of the National e-Infrastructure. CW thanks Jieliang Hong for sharing his knowledge of probability and the authors thank Steve Tobias, Jeffrey Parker,  Andrew Hillier and Zhan Wang for useful discussions. 

For the purpose of open access, the author has applied a Creative Commons Attribution (CC BY) licence to any Author Accepted Manuscript version arising from this submission.

\section*{Data access statement}


The numerical code used to perform the simulations of the full MHD equations and the data of the white noises in the realisations of figures 10, 12 and 14 are available at

https://github.com/Chen-UIC/MHD-zonostrophic-instability
 
 \section*{Declaration of Interests }
 The authors report no conflict of interest.
 
\appendix

\section{The white noise}

In this appendix,  we give a  brief introduction to white noise, which can be found in standard textbooks on stochastic processes or stochastic differential equations, and outline its implementation in our numerical simulations.  We first introduce the Wiener process or Brownian motion $W(t)$: its probability density  at time $t$ given its value $w'$ at $t'<t$ is
\begin{equation}
\rho(w,t|w',t')=\frac{1}{\sqrt{2\pi(t-t')}}\, e^{-\frac{1}{2}\frac{(w-w')^2}{t-t'}}. \label{W1}
\end{equation}
A (real) white noise is then defined as 
\begin{equation}
\xi(t)=\frac{W(t+\Delta t)-W(t)}{\Delta t}, \label{W2}
\end{equation}
in the limit $\Delta t \to 0$.  It has the property 
\begin{equation}
\mathbb{E}[\xi(t_1)\xi(t_2)]=\delta (t_1-t_2),\label{W3}
\end{equation}
i.e.\ white noise at two different times is decorrelated.  While (\ref{W3}) is reached in the limit $\Delta t\rightarrow 0$, 
from a practical point of view,  $\Delta t$ should be much smaller than  any other time scale of the flow.
To obtain a complex white noise, we define
\begin{equation}
\hat{\xi}(t)=\frac{\xi_\mathrm{r}(t)+\mathrm{i}\xi_\mathrm{i}(t)}{\sqrt{2}},\label{W4}
\end{equation}
where $\xi_\mathrm{r}$ and $\xi_\mathrm{i}$ are two  independent real white noises. Expression  (\ref{W4}) is the white noise we use for our forcing (\ref{2.6}) and it satisfies the properties given in (2.7). 
For each numerical realisation, we generate a Brownian motion evaluated at discrete times $t_1, t_2, ..., t_n$ governed by (\ref{W1}),  and then compute the white noise at these time steps via (\ref{W2}) with time step $\Delta t$.   The temporal scheme that we use for our governing equations (2.1) with (2.6) is 
\begin{align}
&\frac{\zeta(t_{n+1})-\zeta({t_n})}{t_{n+1}-t_n}+(-\psi_y\zeta_x+\psi_x\zeta_y+\beta\psi_x)_{t=t_n}\nonumber\\
& \qquad  = \sigma \hat{\xi}(t_n)e^{\mathrm{i}kx}+\mathrm{c.c.}
-\tfrac{1}{2} \mu[{\zeta(t_{n+1})+\zeta(t_n)}]+\tfrac{1}{2}  \nu [ {\nabla^2 \zeta(t_{n+1})+\nabla^2\zeta(t_n)}] .
\end{align}
We use a Crank--Nicolson scheme for the dissipation term to avoid numerical instability.  We evaluate the white noise at the starting time point $t_n$, as per the It\^o interpretation.  We have checked that for a given Brownian motion, reducing our usual time step $\Delta t=0.005$ to $0.0025$ does not change the solution,  and hence the numerical solution is robust.

\section{Comparison with the results of Srinivasan and Young}

In this appendix, we show that our dispersion relation for hydrodynamic zonostrophic instability is identical to \cites{Srinivasan12} result, provided that the same forcing is applied. 
These authors defined a forcing with the property
\begin{equation}
  \mathbb{E}\left[F(x_1,y_1,t_1)F(x_2,y_2,t_2)\right]=\delta (t_2-t_1) \mathcal{F}(x_1-x_2,y_1,y_2).  \label{2.13}
\end{equation}
The $\delta$-function indicates  a rapid decorrelation in time, and the dependence on $x_1-x_2$ indicates that the flow is zonally homogeneous. With the properties of white noise listed in  (\ref{2.6}a--c), our forcing (\ref{2.6}) satisfies (\ref{2.13}), where the spatial structure function is
\begin{equation}\label{2.14}
 \mathcal{F}= 2\sigma^2\cos k \Delta x,
\end{equation}
and $\Delta x=x_1-x_2$.  Its Fourier transform is
\begin{equation}
  \tilde{\mathcal{F}}=\iint \mathcal{F} e^{-\mathrm{i}(p\Delta x+q \Delta y)}\, \mathrm{d}\Delta x\, \mathrm{d}\Delta y=4\pi ^2\sigma^2\left[\delta(p-k)+\delta(p+k)\right]\delta(q).
\end{equation}
To compare to the result of  \citet{Srinivasan12}, we need to consider a variable $a$ defined by $F=\nabla^2 a$, where $a$ is the forcing potential in the corresponding momentum equation. It also has the spatio-temporal correlation function
\begin{align}\label{30}
  &\mathbb{E}[a(x_1,y_1,t_1)a(x_2,y_2,t_2)]=\delta(t_2-t_1)A(x_1-x_2,y_1,y_2).
\end{align}
The Fourier transform of $A$ is
\begin{equation}\label{34}
  \tilde{A}=\frac{\tilde{ \mathcal{F}}}{(p^2+q^2)^2}=\frac{4\pi ^2\sigma^2\left[\delta(p-k)+\delta(p+k)\right]\delta(q)}{(p^2+q^2)^2}\, .
\end{equation}
\citet{Srinivasan12} derived a dispersion relation for a general $\tilde{A}(p,q)$, which is their equation (C16):
\begin{equation}\label{35}
  \bar{s}=\iint\frac{p^2(h_{++}^2-h^2)h^2(h^2-m^2)}{s'h^2h^2_{++}+\mathrm{i}\beta p(h_{++}^2-h^2)}\, \frac{\tilde{A}}{2\mu+2\nu h^2}\, \frac{\mathrm{d}p\,\mathrm{d}q}{(2\pi)^2}\, ,
\end{equation}
where
\begin{align}\label{36}
&\bar{s}=s+\mu+\nu m^2,\quad s'=s+2\mu+\tfrac{1}{2}\nu m^2+2\nu\bigl[p^2+\bigl(q-\tfrac{1}{2} {m}\bigr)^2\bigr], \nonumber \\
&  h=\sqrt{p^2+q^2}\, ,\quad h_{++}=\sqrt{p^2+(q+m)^2}\, .
\end{align}
We note that for $s'$ given by their equation (C12), one should replace $q$ by $q-\tfrac{1}{2} m$ (see the paragraph below their (C15)).  If we substitute (\ref{34}) and (\ref{36}) into (\ref{35}), then after basic algebra, we obtain exactly the same result as our (\ref{2.35}).

\section{Details of MHD zonostrophic instability}
In this appendix, we provide some more details of the analysis of the MHD zonostrophic instability. We first provide some more steps to derive (\ref{3.31}) from (\ref{3.28}) and (\ref{3.29}).   In (\ref{3.28}) and (\ref{3.29}),  the expectations of  quadratic terms of the fundamental waves are derived using a similar method to that used for (\ref{2.31}): we have
\begin{equation}\label{C1}
\left(\begin{array}{llll}
  \mathbb{E}[\hat{\psi}_1^*(t)\hat{\psi}_1(\tau)]\\
   \mathbb{E}[\hat{\psi}_1^*(t)\hat{a}_1(\tau)]\\
    \mathbb{E}[\hat{a}_1^*(t)\hat{\psi}_1(\tau)]\\
     \mathbb{E}[\hat{a}_1^*(t)\hat{a}_1(\tau)]\\
\end{array}
\right)=
\left(\begin{array}{cc}
W_{\psi\psi+} & W_{\psi\psi-}\\
W_{\psi a+}&  W_{\psi a-}\\
W_{a\psi+}&  W_{a \psi-}\\
W_{aa+}&  W_{a a-}
\end{array}
\right)
\left(\begin{array}{ll}
e^{\lambda_{1+}^*(t-\tau)}\\
 e^{\lambda_{1-}^*(t-\tau)}
\end{array}
\right),
\end{equation}
where
\begin{align}
W_{\psi\psi+}=\frac{\Psi_{+}^*\Psi_{-}}{\lambda_{1+}^*+\lambda_{1-}}-\frac{|\Psi_+|^2}{\lambda_{1+}^*+\lambda_{1+}}\, ,\quad 
W_{\psi\psi-}=\frac{\Psi_{+}\Psi_-^*}{\lambda_{1-}^*+\lambda_{1+}}-\frac{|\Psi_{-}|^2}{\lambda_{1-}^*+\lambda_{1-}}\, ,\nonumber
\end{align}
\begin{align}
W_{\psi a+}=\frac{\Psi_{+}^*A}{\lambda_{1+}^*+\lambda_{1-}}-\frac{\Psi_{+}^*A}{\lambda_{1+}^*+\lambda_{1+}}\, ,\quad
W_{\psi a-}=\frac{\Psi_{-}^*A}{\lambda_{1-}^*+\lambda_{1+}}- \frac{\Psi_{-}^*A}{\lambda_{1-}^*+\lambda_{1-}}\, ,\nonumber
\end{align}
\begin{align}\label{C2}
W_{a\psi+}=\frac{A^*\Psi_-}{\lambda_{1+}^*+\lambda_{1-}}-\frac{A^*\Psi_+}{\lambda_{1+}^*+\lambda_{1+}}\, ,\quad
W_{a\psi-}=\frac{A^*\Psi_+ }{\lambda_{1-}^*+\lambda_{1+}}-\frac{A^*\Psi_-}{\lambda_{1-}^*+\lambda_{1-}}\, ,\nonumber 
\end{align}
\begin{align}
W_{aa+}=\frac{|A|^2}{\lambda_{1+}^*+\lambda_{1-}}-\frac{|A|^2}{\lambda_{1+}^*+\lambda_{1+}}\, ,\quad
W_{aa-}=\frac{|A|^2}{\lambda_{1-}^*+\lambda_{1+}}-\frac{|A|^2}{\lambda_{1-}^*+\lambda_{1-}}\, .
\end{align}
We take the expectations of (\ref{3.28}) and (\ref{3.29}), assuming the expectations of the fundamental wave and the mean flow or mean field are separable as in (\ref{assumption2}), then substitute in (\ref{C1}) and (\ref{C2}). After tedious algebra to collect terms with the same exponentials, we arrive at
\begin{align}\label{UB}
&\frac{\mathrm{d}\mathbb{E}[\hat{U}]}{\mathrm{d}t} +\mu\mathbb{E}[\hat{U}]+\nu m^2\mathbb{E}[\hat{U}] \nonumber \\
=\int_0^t&\left[DD_{++}e^{(\lambda_{2+}+\lambda_{1+}^*)(t-\tau)}+DD_{+-}e^{(\lambda_{2+}+\lambda_{1-}^*)(t-\tau)}\right.\nonumber \\
&\quad \left.+DD_{-+}e^{(\lambda_{2-}+\lambda_{1+}^*)(t-\tau)}+DD_{--}e^{(\lambda_{2-}+\lambda_{1-}^*)(t-\tau)}+\mathrm{c.c.}\right]\mathbb{E}[\hat{U}(\tau)]\mathrm{d}\tau \nonumber \\
+\int_0^t&\left[DM_{++}e^{(\lambda_{2+}+\lambda_{1+}^*)(t-\tau)}+DM_{+-}e^{(\lambda_{2+}+\lambda_{1-}^*)(t-\tau)}\right.\nonumber \\
&\quad \left.+DM_{-+}e^{(\lambda_{2-}+\lambda_{1+}^*)(t-\tau)}+DM_{--}e^{(\lambda_{2-}+\lambda_{1-}^*)(t-\tau)}+\mathrm{c.c.}\right]\mathbb{E}[\hat{B}(\tau)]\mathrm{d}\tau,
\end{align}
\begin{align}
\label{BU}
& \frac{\mathrm{d}\mathbb{E}[\hat{B}]}{\mathrm{d}t}  +\eta m^2\mathbb{E}[\hat{B}] \nonumber \\
=\int_0^t &\left[MD_{++}e^{(\lambda_{2+}+\lambda_{1+}^*)(t-\tau)}+MD_{+-}e^{(\lambda_{2+}+\lambda_{1-}^*)(t-\tau)}\right.\nonumber \\
&\quad \left.+MD_{-+}e^{(\lambda_{2-}+\lambda_{1+}^*)(t-\tau)}+MD_{--}e^{(\lambda_{2-}+\lambda_{1-}^*)(t-\tau)}+\mathrm{c.c.}\right]\mathbb{E}[\hat{U}(\tau)]\mathrm{d}\tau \nonumber \\
+\int_0^t& \left[MM_{++}e^{(\lambda_{2+}+\lambda_{1+}^*)(t-\tau)}+MM_{+-}e^{(\lambda_{2+}+\lambda_{1-}^*)(t-\tau)}\right.\nonumber \\
&\quad \left.+MM_{-+}e^{(\lambda_{2-}+\lambda_{1+}^*)(t-\tau)}+MM_{--}e^{(\lambda_{2-}+\lambda_{1-}^*)(t-\tau)}+\mathrm{c.c.}\right]\mathbb{E}[\hat{B}(\tau)]\mathrm{d}\tau,
\end{align}
where $DD_{++}$, $DD_{+-}$ etc. are constants with expressions
\begin{align}
 & DD_{++}=\mathrm{i} m^2 k(-\Lambda D_+ W_{\psi\psi+}+MW_{\psi a+}-\Lambda M W_{a\psi+}-D_-W_{aa+}),\nonumber \\
 & DD_{+-}=\mathrm{i} m^2 k(-\Lambda D_+ W_{\psi\psi-}+MW_{\psi a-}-\Lambda M W_{a\psi-}-D_-W_{aa-}), \nonumber \\
 & DD_{-+}=\mathrm{i} m^2 k(+\Lambda D_- W_{\psi\psi+}-MW_{\psi a+}+\Lambda M W_{a\psi+}+D_+W_{aa+}),  \nonumber \\
 & DD_{--}=\mathrm{i} m^2 k(+\Lambda D_- W_{\psi\psi-}-MW_{\psi a-}+\Lambda M W_{a\psi-}+D_+W_{aa-}),   \label{B5}
\end{align}
\begin{align}
& DM_{++}=\mathrm{i} m^2k(-MW_{\psi\psi+}+\Lambda D_+W_{\psi a+}+D_-W_{a\psi+}+\Lambda M W_{aa+}),\nonumber \\
& DM_{+-}=\mathrm{i} m^2k(-MW_{\psi\psi-}+\Lambda D_+W_{\psi a-}+D_-W_{a\psi-}+\Lambda M W_{aa-}),\nonumber \\
& DM_{-+}=\mathrm{i} m^2k(+MW_{\psi\psi+}-\Lambda D_-W_{\psi a+}-D_+W_{a\psi+}-\Lambda M W_{aa+}),\nonumber \\
& DM_{--}=\mathrm{i} m^2k(+MW_{\psi\psi-}-\Lambda D_-W_{\psi a-}-D_+W_{a\psi-}-\Lambda M W_{aa-}),\label{B6}
\end{align}
\begin{align}
&MD_{++}=\mathrm{i}m^2 k(+\Lambda M W_{\psi \psi+}+D_-W_{\psi a+}+\Lambda D_+W_{a \psi+}-M W_{aa+}), \nonumber \\
&MD_{+-}=\mathrm{i}m^2 k(+\Lambda M W_{\psi \psi-}+D_-W_{\psi a-}+\Lambda D_+W_{a \psi-}-M W_{aa-}), \nonumber \\
&MD_{-+}=\mathrm{i}m^2 k(-\Lambda M W_{\psi \psi+}-D_+W_{\psi a+}-\Lambda D_-W_{a \psi+}+M W_{aa+}),\nonumber \\
&MD_{--}=\mathrm{i}m^2 k(-\Lambda M W_{\psi \psi-}-D_+W_{\psi a-}-\Lambda D_-W_{a \psi-}+M W_{aa-}),\label{B7}
\end{align}
\begin{align}
& MM_{++}=\mathrm{i}m^2 k(-D_-W_{\psi\psi+}-\Lambda MW_{\psi a+}+MW_{a\psi+}-\Lambda D_+ W_{aa+}), \nonumber \\
& MM_{+-}=\mathrm{i}m^2 k(-D_-W_{\psi\psi-}-\Lambda MW_{\psi a-}+MW_{a\psi-}-\Lambda D_+ W_{aa-}), \nonumber \\
& MM_{-+}=\mathrm{i}m^2 k(+D_+W_{\psi\psi+}+\Lambda MW_{\psi a+}-MW_{a\psi+}+\Lambda D_- W_{aa+}), \nonumber \\
& MM_{--}=\mathrm{i}m^2 k(+D_+W_{\psi\psi-}+\Lambda MW_{\psi a-}-MW_{a\psi-}+\Lambda D_- W_{aa-}). \label{B8}
\end{align}
Applying the exponential  form (\ref{3.30}) to (\ref{UB}) and (\ref{BU}), we obtain (\ref{3.31}) with constants
\begin{align}
N_{UU}=\frac{DD_{++}}{s-\lambda_{2+}-\lambda_{1+}^*}+\frac{DD_{+-}}{s-\lambda_{2+}-\lambda_{1-}^*}+\frac{DD_{-+}}{s-\lambda_{2-}-\lambda_{1+}^*}+\frac{DD_{--}}{s-\lambda_{2-}-\lambda_{1-}^*}+\mathrm{c.c.e.}s., \nonumber \\
N_{UB}=\frac{DM_{++}}{s-\lambda_{2+}-\lambda_{1+}^*}+\frac{DM_{+-}}{s-\lambda_{2+}-\lambda_{1-}^*}+\frac{DM_{-+}}{s-\lambda_{2-}-\lambda_{1+}^*}+\frac{DM_{--}}{s-\lambda_{2-}-\lambda_{1-}^*}+\mathrm{c.c.e.}s., \nonumber \\
N_{BU}=\frac{MD_{++}}{s-\lambda_{2+}-\lambda_{1+}^*}+\frac{MD_{+-}}{s-\lambda_{2+}-\lambda_{1-}^*}+\frac{MD_{-+}}{s-\lambda_{2-}-\lambda_{1+}^*}+\frac{MD_{--}}{s-\lambda_{2-}-\lambda_{1-}^*}+\mathrm{c.c.e.}s.,  \nonumber \\
N_{BB}=\frac{MM_{++}}{s-\lambda_{2+}-\lambda_{1+}^*}+\frac{MM_{+-}}{s-\lambda_{2+}-\lambda_{1-}^*}+\frac{MM_{-+}}{s-\lambda_{2-}-\lambda_{1+}^*}+\frac{MM_{--}}{s-\lambda_{2-}-\lambda_{1-}^*}+\mathrm{c.c.e.}s. \label{Ns}
\end{align}
When $B_0=0$, only $DD_{++}$ and $MM_{-+}$ are not zero, which corresponds to the two dispersion relations in (\ref{3.33}).

Next, we give the main steps used to derive the simplified dispersion relations in the parameter limits discussed in \S 3.4. For convenience, we use the parameters before rescaling.  
  In the limit of large $\eta$
\begin{equation}
\lambda_{1-}\sim -\eta k^2, \quad \lambda_{2-}\sim -\eta (k^2+m^2)
\end{equation}
grow very large, in contrast to $\lambda_{1+}$ and $\lambda_{2+}$ that remain bounded (cf.\  (\ref{lambda})).   Hence in (\ref{Ns}),  all terms with $\lambda_{1-}$ and $\lambda_{2-}$ become negligible.  A more detailed analysis indicates that $DM_{++}$,  $MD_{++}\sim \eta^{-1}$ and so are also small,  making the coupling terms $N_{BU}$ and $N_{UB}$ negligible. Hence the only leading order term left is the $DD_{++}$ term.  
 
 In the limit of small $\eta$ and $B_0$, the derivation of (\ref{3.41}) requires careful analysis of the orders of various terms in (\ref{Ns}).  The outcome is that only the $DD_{--}\sim O(B_0^2/\eta)$ in  $N_{UU}$ remains at leading order. The contribution from other terms is either at $O(B_0^2)$ or $O(B_0^4/\eta)$ or smaller. 

In the limit of small $\eta$ and large $B_0$, while $N_{BU}$ and $N_{UB}$ are at $O(B_0^{-1})$, $N_{UU}$ and $N_{BB}$ are at $O(B_{0}^{-2})$ and thus drop out of  the leading order terms of (\ref{3.32}), given that $s$ is also small. Equations (\ref{3.44}--\ref{3.46}) correspond to $\eta=0$; including  small $\eta$ only adds $O(\eta/B_0)$ corrections to $N_{UU}$ and $N_{BB}$ which are of order $O(1/B_0)$. We have used the software Maplesoft\textsuperscript \textregistered to derive the asymptotic dispersion relation in this limit. 



\bigskip

\bibliographystyle{jfm}
\bibliography{jfm-instructions}

\end{document}